\documentclass[a4paper,11pt]{article}
\usepackage{jheppub}
\usepackage[utf8]{inputenc}
\usepackage{url}
\usepackage{mathrsfs}  
\usepackage{pdflscape}

\usepackage{refcount}
\usepackage{booktabs}
\usepackage{todonotes}
\usepackage{enumitem}
\usepackage{adjustbox}
\usepackage{afterpage}
\usepackage{dsfont}
\usepackage{nameref}
\usepackage{dsfont}
\usepackage{makecell}
\usetikzlibrary{fit}
\usepackage{longtable}
\usepackage{float}
\usepackage{mdframed}

\usetikzlibrary{decorations.markings}

\makeatletter
\newcommand\level[1]{%
  \ifcase#1\relax\expandafter\chapter\or
    \expandafter\section\or
    \expandafter\subsection\or
    \expandafter\subsubsection\else
    \def\next{\@level{#1}}\expandafter\next
  \fi}
\newcommand{\@level}[1]{%
  \@startsection{level#1}
    {#1}
    {\z@}%
    {-3.25ex\@plus -1ex \@minus -.2ex}%
    {1.5ex \@plus .2ex}%
    {\normalfont\normalsize\bfseries}}

\newcounter{level4}[subsubsection]
\@namedef{thelevel4}{\thesubsubsection.\arabic{level4}}
\@namedef{level4mark}#1{}
\count@=4
\loop\ifnum\count@<100
  \begingroup\edef\x{\endgroup
    \noexpand\newcounter{level\number\numexpr\count@+1\relax}[level\number\count@]
    \noexpand\@namedef{thelevel\number\numexpr\count@+1\relax}{%
      \noexpand\@nameuse{thelevel\number\count@}.\noexpand\arabic{level\number\numexpr\count@+1\relax}}
    \noexpand\@namedef{level\number\numexpr\count@+1\relax mark}####1{}}
  \x
  \advance\count@\@ne
\repeat
\makeatother
\usepackage{tikz}
\usetikzlibrary{decorations.pathreplacing, positioning}

\usepackage{pdflscape}

\usepackage[utf8]{inputenc} 
\usepackage{amsmath}
\usepackage{amsthm}
\usepackage{amsfonts,mathrsfs} 
\usepackage{xcolor}
\usepackage[english]{babel} 
\usepackage[autostyle]{csquotes}
\usepackage{mathtools}

\usepackage{slashed}

\newlist{inparaenum}{enumerate}{3}
\setlist[inparaenum,1]{label=\arabic*.}
\setlist[inparaenum,2]{label=\emph{\alph*})}
\setlist[inparaenum,3]{label=\emph{\roman*})}

\numberwithin{equation}{section}

\makeatletter
\newcommand\footnoteref[1]{\protected@xdef\@thefnmark{\ref{#1}}\@footnotemark}
\makeatother

\preprint{Imperial/TP/25/AH/10}

\title{Chiral ring along the RG flow in 5d $\mathcal{N}=1$}

\author{Amihay Hanany}
\author{and Elias Van den Driessche}
\affiliation{Theoretical Physics Group, Blackett Laboratory, Imperial College London, Prince Consort Road
London, SW7 2AZ, UK}
\emailAdd{a.hanany@imperial.ac.uk}
\emailAdd{e.van-den-driessche24@imperial.ac.uk}

\begin{document}

\abstract{We compute the Higgs branch chiral ring of a simple class of 5d theories at strong coupling. A deformation by the instanton mass implies that the chiral ring at weak coupling is corrected by a nilpotent operator, the gaugino bilinear. Consequently, F-term equations alone do not suffice to completely determine the moduli space of vacua and perturbative non-renormalization arguments are invalidated.}

\maketitle

\section{Introduction}
5d $\mathcal{N}=1$ supersymmetric gauge theories flow between two RG fixed points. The first one, in the IR, corresponds to the free theory. The second one, in the UV, corresponds to the strongly coupled theory, since the squared coupling constant has inverse mass dimension:
\begin{equation}
    [m]=\bigg[\dfrac{1}{g^2}\bigg]\ .
     \label{eq:mass}
\end{equation}
The existence of this second fixed point was not obvious, since the theory is power counting non renormalizable, and was argued for in \cite{seiberg1996five, morrison1997extremal, intriligator1997five} initially for the case of $SU(2)$ gauge theories. Asymptotic safety is not always guaranteed: depending on the number and representations of the flavours, the fixed point can exist in 5d, in 6d \cite{jefferson2018geometric, jefferson2023towards} or does not exist at all \cite{kim2015tao}. In the case of SU(2), a 5d interacting fixed point is allowed for up to 7 flavours in the fundamental representation. Similar conditions arise for different gauge groups. \\

\noindent Starting from a weakly coupled description of the theory, we can build a current out of the field strength $F$:
\begin{equation}
    j= \star \text{tr}F \wedge F \ ,
    \label{eq:current}
\end{equation}
with $\star$ representing Hodge duality and $\text{tr}$ the trace over the adjoint representation of the gauge group. Such current is topologically conserved due to the Bianchi identity and corresponds to a U(1) symmetry. The operators carrying U(1) charge are called instantons, defined as 1/2 BPS local disorder operators in the path integral formulation of the theory. In 5d, gauge instantons are not codimension 5, but have one extended direction. As was shown in the original papers, instantons are massless at the 5d interacting fixed point, if existing. Switching on their mass, i.e. integrating them out, activates an RG flow to finite, weak and ultimately zero coupling. The instantons are thus an inherently non perturbative effect and they can lead to flavour symmetry enhancement, for special combinations of colours and flavours. Instanton operators in 5d have been the subject of many studies, including \cite{tachikawa2015instanton, rodriguez2015supersymmetrizing, zafrir2015instanton, yonekura2015instanton, hanany2025higgs, benetti2024comments, bertolini2025symmetry,lambert2015instanton}. \\

\noindent All along the RG flow, including at the interacting fixed point, 5d $\mathcal{N}=1$ theories have a two-branched moduli space of vacua, a Coulomb branch and a Higgs branch. Conformal and flavour symmetry are spontaneously broken on a generic point of either branch and the gauge symmetry is Higgsed to some subgroup (or completely Higgsed, on the Higgs branch). The low energy theory on the Higgs branch is an interacting theory of mesons, gaugino bilinear and instantons. Mesons and gaugino bilinear are the moment maps of, respectively, the flavour symmetry and the topological U(1) symmetry. Hence, for the pure theory there will be no mesons but there will always be a gaugino bilinear, regardless of the flavours. If we integrate the instantons out of the abovementioned low energy theory, we obtain some interacting theory of only mesons and gaugino bilinear at finite coupling.\\

\begin{figure}[H]
    \centering
\begin{tikzpicture}[>=stealth, thick]

\node[circle, fill=black, inner sep=2pt, label={[align=center, above] 
  Gaugino bilinear,\\ instantons and mesons \\ at strong coupling}] (A) at (0,4) {};

\node[circle, fill=black, inner sep=2pt, 
      label={[align=center, below, yshift=-6pt] free mesons}] (B) at (10,0) {};

\node[anchor=east, font=\bfseries] at (-1,2) {\boxed{\textbf{UV}}};

\node[anchor=west, font=\bfseries] at (11,2) {\boxed{\textbf{IR}}};

\draw[->,postaction={decorate}, decoration={markings, mark=at position 0.55 with {\arrow{>}}}]
  (A) .. controls (1,-1) and (9,4) .. (B)
  node[midway, above, yshift=2pt, align=center] {\textit{integrating out the instantons}}
  node[midway, below, yshift=-2pt, align=center] {Gaugino bilinear, mesons \\ at weak coupling};

\end{tikzpicture}
    \caption{RG flow activated by the instanton mass.}
    \label{fig:RGflow}
\end{figure}
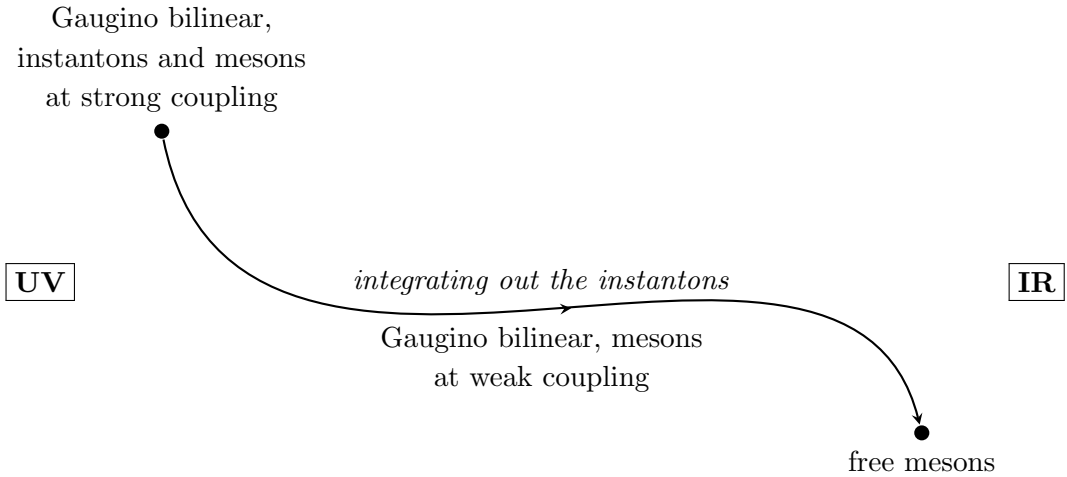

\noindent We will focus on the scalar fields of the low energy theory which are 1/2 BPS local gauge invariant operators. They are not independent of each other, but rather they satisfy algebraic constraints, which determine the geometry of the moduli space of vacua. While some of these constraints were worked out for the theory at finite coupling, there has been no work yet on the theory in the UV limit, except for SU(2) in \cite{cremonesi2017instanton}. In the present work we will verify that the theory at the strong coupling fixed point has indeed mesons, gaugino bilinear and instantons, and we will work out how and why these scalar fields constrain one another.\\

\noindent A first nontrivial result of the present work is thus the full set of such constraints for the superconformal theory at the infinite coupling fixed point. These constraints concern $Sp(k)$ theories\footnote{We follow the convention according to which $Sp(k)$ has rank $k$ and its fundamental representation is $2k$ dimensional. Its dual Coxeter number is $h^{\vee}=k+1$.} with $N_f \leq 2k+5$ flavours in the fundamental representations. These theories are a most natural starting point to study Higgs branches, since they have no Chern-Simons levels (unlike SU theories) and no baryons (unlike SO and SU). They have also been recently studied in \cite{hanany2025higgs}, in the context of dressed instanton operators.\\

\noindent It has been supposed that the low energy theory at finite coupling only includes the mesons, without any mention of the gaugino bilinear. Nonetheless, we know that there exist a topological U(1) symmetry, hence we expect also its moment map. In the present work we integrate out the instantons and flow to finite and weak coupling, finding that the low energy theory includes the gaugino bilinear $S$. In particular, at finite coupling, $S$ is nilpotent; the following equations holds:
\[
S^{h^{\vee}}=0 \qquad \text{Nilpotency of gaugino bilinear at finite coupling} \ ,
\]
which implies that only a finite number of operators involving $S$ in nonvanishing.\\

\noindent We will determine the constraints between the mesons and $S$ at finite coupling and find important corrections to the current understanding. For example, we consistently find:
\[
\text{tr}M^2=S^2
\]
for $k>1$ and any flavour. This implies that the Higgs branch at finite coupling of $Sp(k)$, with $k>1$, is not the closure of a nilpotent orbit of $D_{N_f}$, as widely believed. This also implies that the F-term relations are not taking into account $S$ and are thus describing an incomplete moduli space of vacua, as already pointed out in \cite{bourget2020brane}.\\

\noindent A further consequence is the following. Since \cite{seiberg1993naturalness, argyres1996moduli}, it is believed that the Higgs branch does not get quantum corrections, since the gauge coupling belongs to a background vector multiplet. We find instead that the Higgs branch gets contributions all along the RG flow due to the gaugino bilinear (at finite coupling) and the instantons (at infinite coupling). While the latter effect was already identified in \cite{seiberg1996five}, the correction at finite coupling appears here for the first time and can be potentially recovered via a weak coupling computation, although we will not present it here. It would be interesting to understand exactly which points of the non-renormalization arguments are invalidated by the presence of a nilpotent operator.\\

\noindent The structure of the paper follows the RG flow from strong coupling to zero coupling: in Sec. \ref{sec:generators} we will present and explain the low energy scalars at strong, finite and zero coupling. In Sec. \ref{sec:constraints} we will explain which kind of constraints define the Higgs branch. Finally, in Sec. \ref{sec:methods} we will present some explicit cases to get a better understanding of how the constraints work and how we found them.\\

\section{Generators of the chiral ring along the RG flow}\label{sec:generators}
The chiral ring is the set of gauge invariant local chiral operators, which by definition are annihilated by half of the supercharges. Since any product or sum of chiral operators is still chiral, they form a ring. In theories with 8 real supercharges, like 5d $\mathcal{N}=1$, there are two branches of the moduli space of vacua, Higgs and Coulomb branch. We will only deal with the Higgs branch, as it has received comparatively less attention than the Coulomb branch, partly due to the belief that it is non-renormalized under RG flow.  We assume an equivalence between holomorphic functions on the moduli space of vacua and chiral operators in the chiral ring. We are going to show that the chiral ring is corrected all along the RG flow activated by the instanton mass.
\subsection*{Infinite coupling}
 The generators of the ring at infinite coupling are the following:
\begin{itemize}
    \item Mesons $M$\\
The mesons are the moment map of the flavour symmetry, i.e. they are the superconformal primary in the “conserved current" superconformal multiplet\footnote{The notations for 5d conformal representation is: $[a,b]^{(c)}_d$, with $(c)$ highest weight of $SU(2)_R$ representation, $d$ scaling dimension and $[a,b]$ SO(5) representation, such that, for example, $[1,0]$ is a vector and $[0,1]$ is a spinor.}:
\begin{equation}
    [0,0]_3^{(2)}\to[0,1]_{7/2}^{(1)}\to [0,0]_4^{(0)}+[1,0]_4^{(0)}
    \label{eq:shortmultiplet}
\end{equation}
\noindent whose elements all transform in the adjoint representation of the flavour symmetry. At strong coupling, quarks (i.e. the flavour hypermultiplets) need not exist as fundamental operators, so mesons should be viewed as genuine generators rather than composites.\\If the gauge group is $Sp(k)$, as in the rest of the paper, the flavour symmetry is $SO(2N_f)$, hence $M$ will be an antisymmetric matrix $M_{[ij]}$.
The component $[1,0]^{(0)}_4$ is the $SO(2N_f)$ flavour current and $[0,0]^{(0)}_4$ is the scalar that couples to the masses of the flavours and deforms the theory.
    \item Gaugino bilinear $S$\\
The gaugino bilinear $S$ is the moment map of the topological $U(1)$ instanton number symmetry. It belongs to the same type of short multiplet as in Eq. \ref{eq:shortmultiplet}, but since the adjoint of $U(1)$ is trivial, these components are flavor singlets. The component $[1,0]^{(0)}_4$ is the topological current, while $[0,0]_4^{(0)}$ is the scalar that couples \cite{tachikawa2015instanton} to the deformation parameter $m_0=1/g^2$, namely the instanton mass.
    \item Instantons $I,\tilde{I}$ \\
Instantons are local disorder operators, namely they are defined as constraints in the path integral over the gauge degrees of freedom of the theory. In particular, they constrain the gauge field strength to satisfy (up to a normalization of the integrand):
\[
\int \star \text{tr}F \wedge F= \text{ }n \in \mathbb{Z} 
\]
where $n$ is the instanton charge and the integrand corresponds to the U(1) conserved current.\\
\noindent Instantons act on states by modifying their instanton number, thus allowing to move between different topological sectors. They are known to have the following scaling dimensions and R-charge, respectively \cite{cremonesi2017instanton, tachikawa2015instanton, kim20125}:
\begin{equation}
    \begin{array}{ll}
        \Delta_I & = 3h^{\vee}/2 \\
        R_I & = h^{\vee}
    \end{array}
\end{equation}
hence they can be identified as living in the short multiplet \cite{cordova2019multiplets}:
\[
\quad [0,0]^{(R)}_{3R/2} \to [0,1]^{(R-1)}_{3R/2+1/2} \to [1,0]^{(R-2)}_{3R/2+1}+[0,0]^{(R-2)}_{3R/2+1}\to [0,1]^{(R-3)}_{3R/2+3/2} \to [0,0]^{(R-4)}_{3R/2+2} \ .
\]
Let us notice that such multiplet reduces to the free hypermultiplet in the case of $R=1$, and has the same form of the conserved current multiplet in Eq. \ref{eq:shortmultiplet} for $R=2$ (i.e. $Sp(1)$). \\

\noindent The instanton and anti-instanton have opposite U(1) charge, which we normalize to 1 and $-1$. For $SO(2N_f)$ flavour symmetry, they transform in the spinor representations. In particular for odd $N_f$ they transform in the two different spinor representations, being the conjugate of each other, while for even $N_f$ in the same one.
\end{itemize}

\subsection*{Weak coupling}
As identified since \cite{seiberg1996five}, the instanton mass corresponds the inverse square of the gauge coupling constant. Hence, flowing from strong coupling to weak coupling means switching on the instanton mass and decoupling them from the chiral ring. Consequently, the only remaining generators are the moment maps of the symmetries.
\begin{itemize}
    \item Mesons\\
     The multiplet\footnote{The supermultiplet notation at finite coupling is different from the conformal case: fields are denoted as $[a]_c^{(b)}$, where $[a]$ is a representation of the little group SO(3) in 5d, $(b)$ is the highest weight of the representation of the R-symmetry $SU(2)_R$, $c$ is the scaling dimension.} in which the mesons transform is:
\begin{equation}
    [0]_3^{(2)}\to[1]_{7/2}^{(1)}\to [2]_4^{(0)} \ .
\end{equation}
 Such multiplet can also be seen as the symmetric product of two hypermultiplets.
Indeed, from a IR point of view, we know the mesons to be defined as:
\begin{equation}
    M_{[ij]}=Q_i^aQ^b_j\Omega_{[ab]}
    \label{eq:mesonsfinite}
\end{equation}
\noindent with: $i$ are flavour indices, $a$ are gauge indices, $Q$ represent the flavours in the vector representation of the gauge group and $\Omega_{ab}$ the symplectic invariant. Given Eq. \ref{eq:mesonsfinite}, it is obvious that at finite coupling the mesons transform as the primary representation in the symmetric product of the hypermultiplets.\\

\item Gaugino bilinear $S$\\
We propose a weak coupling definition of the gaugino bilinear, in analogy with the definition of the glueball in 4d. We take a bilinear in the gaugino and multiply by the appropriate power of the coupling constant to ensure the right scaling dimensions:
\[
S^{(\alpha\beta)}=g^2 \text{Tr}\bar{\lambda}^{\alpha}\lambda^{\beta} \ ,
\]
where $\bar{\lambda}=\lambda^TC$, with $C$ the charge conjugation matrix, $\alpha,\beta$ $SU(2)_R$ indices. The gaugino bilinear $S$ which is going to appear in the rest of the paper corresponds to the highest charge component of $S^{\alpha\beta}$ under $SU(2)_R$.
\noindent In the simple case of gauge group SU(2), the gaugino bilinear is $\lambda_{(ab)}$ with $a,b=1,2$ and:
\[
S^{(\alpha\beta)}=g^2\big[\bar{\lambda}_{11}^{\alpha}\cdot \lambda_{22}^{\beta}-\bar{\lambda}_{12}^{\alpha}\cdot \lambda_{12}^{\beta}\big] \qquad \text{gaugino bilinear for SU(2)} \ .
\]

\noindent The spinor bilinear is symmetric \cite{freedman2012supergravity} in the $SU(2)_R$ indices. The trace is over the colour degrees of freedom, as the gaugino transforms in the adjoint representation of the gauge group. The supermultiplet of the gaugino bilinear at weak coupling is then part of the symmetric product of two vectormultiplets.\\

\noindent The gaugino bilinear is a composite operator made of fermionic fields and it will vanish if raised to an appropriate power (namely equal to the dimension of the adjoint representation of the gauge group plus one). However, as found also in 4d $\mathcal{N}=1$ in \cite{cachazo2003chiral, witten2003chiral}, it vanishes much earlier than that. Indeed, at weak coupling, we have that $S^{h^{\vee}}=0$, with $h^{\vee}$ the dual Coxeter number of the gauge group. Furthermore, we argue that we cannot apply factorization and cluster decomposition in the case of $S$ in the chiral ring. Being $S$ an operator made out of fermions, unlike the other generators, there is no reason why in the chiral ring $S^{h^{\vee}}=0$ should imply $S^a=0$, with $a < h^{\vee}$. Being nilpotent, the gaugino bilinear contributes only a discrete set of chiral ring operators, but it still affects the chiral ring relations.
\end{itemize}

\subsection*{Zero coupling}
At the other end of the RG flow, at zero coupling, we only have $N_f$ free hypermultiplets. Hence there is no U(1) topological symmetry and there is no gaugino bilinear either. Consequently, the chiral ring is generated solely by the mesons, satisfying the constraints specified in the next section.

\section{Constraints along the RG flow}\label{sec:constraints}
At no point in the RG flow is the Higgs branch freely generated by the above described generators. On the contrary, relations among them characterize geometrically the moduli space of vacua, as described in 4d $\mathcal{N}=1$ in \cite{intriligator1995phases, cachazo2003chiral, witten2003chiral}. Such relations are exact algebraic relations, hence they are valid on all states of the theory.\\

\noindent Since chiral operators form a ring, their operator product expansion trivializes: inside the chiral ring all OPE coefficients will be either zero or one, with no singular terms. In this perspective, the relations that we are about to describe can be seen as OPE expansions, defining the algebraic structure of the Higgs branch.\\

\noindent The following relations have been tested systematically for $Sp(2)$ with $N_f=1\dots9$, and for $SU(2)$ in \cite{cremonesi2017instanton}, as well as in some sporadic cases at higher rank. We conjecture them to appear in general for $Sp(k)$ with any allowed number of flavours. Let us indeed recall that, at infinite coupling, $Sp(k)$ gauge theories only allow up to $N_f=2k+5$ flavours. Such constraint \cite{jefferson2018geometric, kim2015tao, bergman20155d, bourget2020magnetic} can arise in two equivalent ways: either from requiring non intersection of external legs in the braneweb construction of the theory, or from requiring non-negative curvature on the Coulomb branch.

\subsection*{Infinite coupling}
\begin{itemize}
    \item \textbf{Mesons and gaugino bilinear}\\
    The mesons and the gaugino bilinear are constrained by:
\begin{equation}
    \boxed{M^2_{ij}=S^2\delta_{ij}}
    \label{eq:casimireq}
\end{equation}
such relation transforms in a reducible representation of $SO(2N_f)$, namely the 2nd rank symmetric and the singlet. Taking the trace, we isolate the singlet representation:
\begin{equation}
    \text{tr}M^2=S^2 \ ,
    \label{eq:trM^2=S^2}
\end{equation}
up to some coefficient, equal to twice $N_f$, which can be reabsorbed in the definition of the generators. Given $N_f$ flavours, we can construct $N_f$ Casimir invariants of the flavour symmetry, namely:
\begin{equation}
    \text{trM}^2\ , \text{tr}M^4 \ , \dots , \text{tr}M^{2a} \quad a=1 \dots N_f-1
    \label{eq:higherordertraces}
\end{equation}
and
\begin{equation}
    \text{Pf}M=\epsilon_{i_1\dots i_{N_f}}M_{i_1i_1}\cdots M_{i_{N_f-1}i_{N_f}} \ .
\end{equation}
In particular, Eq. \ref{eq:casimireq} implies that all the Casimir invariants of the type \ref{eq:higherordertraces} get corrected by appropriate powers of the gaugino bilinear. In particular, a correction of the kind:
\[
\text{tr}M^{2k}=S^{2k}
\]
corresponds to the highest weight of a spin-$k$ representation of the R-symmetry, hence it implies the existence of other $2k+1$ relations, involving non holomorphic fields.\\

\noindent Additional relations among $M_{ij}$ and $S$ may also be present, involving antisymmetric product of the mesons.  For example, for $N_f=3$ $Sp(2)$:
\[
\epsilon_{ijklmn}M_{kl}M_{mn}=SM_{ij} \ ,
\]
however these are low rank special cases, appearing only for $N_f \leq 2k+1$. Indeed what happens is that, for general $Sp(k)$, at R-charge $2p$ we can form the antisymmetrized product of $p$ mesons. Such operator can be corrected by $S^a$ times a lower number $b$ of mesons, as long as the latter representation is Hodge dual to the former one. Hence there exist corrections of the type\footnote{We use the wedge symbol for ease of notation, for example: $\wedge^2M_{ij}=M_{[ij}M_{kl]}$. The epsilon symbol is the completely antisymmetric invariant of $SO(2N_f)$.}:
\begin{equation}
    \boxed{\epsilon\cdot \wedge^pM=S^a \, (\wedge^bM)}
    \label{eq:Hodgedual}
\end{equation}
with $N_f=p+b$ and $p=a+b$. The notation $\epsilon \cdot$ represents the contraction with the Levi-Civita symbol. If we take maximum order, i.e. $p=k+1$, we find that this correction indeed takes place only for $N_f\leq 2k+1$.

\item \textbf{Mesons, gaugino bilinear and instantons}: lowest R-charge\\
\noindent Let us turn our attention to the relations involving the instantons. From the point of view of $SO(2N_f)$ representation and $U(1)$ charge, the simplest product of generators is the product of a gaugino bilinear and an instanton, which will transform in the spin-$(k+3)$ of the R-symmetry. Any correction to this product will have to match its R-charge, which is the case for the product of the mesons and the instantons, with vector and spinor indices appropriately contracted with a gamma matrix $\gamma$.  We find indeed that these two product operators are constrained:
\begin{equation}
    \boxed{(M\gamma_{})\cdot I=SI} \ ,
    \label{eq:eigenvalue1}
\end{equation}
and same for the anti-instanton, which we denote with a tilde:
\begin{equation}
    \boxed{(M\gamma)\cdot \tilde{I}=S\tilde{I}} \ ,
    \label{eq:eigenvalue2}
\end{equation}
These relations have, respectively, charge $+1$ and $-1$ under the $U(1)_I$, and transform in a reducible representation of the flavour symmetry: a gravitino-like representation (product of vector and spinor representations) and the spinor representation in which $I$ and $\tilde{I}$ transform. \\
Let us notice that Eq. \ref{eq:eigenvalue1} and Eq. \ref{eq:eigenvalue2} resemble eigenvalue equations, with $S$ the eigenvalue of $M$ corresponding to eigenvector $I$.
\item \textbf{Mesons, gaugino bilinear and instantons}: higher R-charge \\
In the previous paragraph we considered relations involving only one instanton at a time. We will now consider instanton bilinears and see how they affect $S$ and $M_{ij}$. 
Instantons have R-charge equal to the dual coxeter number, hence instanton bilinears are going to correct relations at double that R-charge. Any product operator transforming at lower R-charge will not be constrained by instanton bilinears.\\  As mentioned before, instanton and anti-instanton transform in the spinor representation of $SO(2N_f)$. In the case of $N_f$ even, they transform in the same spinor type, while for $N_f$ odd they have opposite spinor type. The tensor product of these representations\footnote{For ease of notation we have replaced the Dynkin labels with the highest weights to define representations: $[n_1,n_2,\dots, n_r]\to\mu_1^{n_1}\mu_2^{n_2}\cdots\mu_r^{nr}$. Hence, for example, the fundamental of $SO(2n)$ would be $\mu_1$ (i.e. $[1,0,\dots,0]$) and the adjoint $\mu_2$ (i.e. $[0,1,0,\dots,0]$). The second rank traceless symmetric would be $\mu_1^2$, i.e. $[2,0,\dots,0]$. We will adopt this notation for all representations of $SO(2N_f)$.} gives:
\begin{equation}
    I \otimes \tilde{I} =
\begin{cases}
\mu_{N_f}\otimes\mu_{N_f-1}=\mu_{N_f}\mu_{N_f-1}+\mu_{N_f-3}+\mu_{N_f-5}+\dots+\mu_2+1, & N_f \ \text{odd}, \\
\mu_{N_f}\otimes \mu_{Nf}=\mu_{N_f}^2+\mu_{N_f-2}+\mu_{N_f-4}+\dots +\mu_2+1, & N_f \ \text{even}.
\end{cases}
\label{eq:spinorbilinear}
\end{equation}
All representation on the r.h.s. in Eq. \ref{eq:spinorbilinear} are corrected by appropriate powers of the gaugino bilinear and antisymmetrized mesons. The general constraint is:
\begin{equation}
  \boxed{ S^c \wedge ^dM_{ij}=(I\cdot\tilde{I)}|_{\mu_{2d}}}
  \label{eq:generalcorrection}
\end{equation}
with $c+d=h^{\vee}$. Notice that some of the elements on the l.h.s. of \ref{eq:generalcorrection} might be also appearing on the r.h.s of \ref{eq:Hodgedual}, in which case you would have an equality between three different operators. Furthermore, if we take $d=0$ in Eq. \ref{eq:generalcorrection}, we obtain:
\begin{equation}
    S^{h^{\vee}}=I\cdot \tilde{I} \ ,
\end{equation}
which transforms in the singlet representation and is always present among the constraints at infinite coupling, regardless of the amount of flavours, if any. In particular it reproduces the previous result of \cite{cremonesi2017instanton} and implies that the Higgs branch at infinite coupling is a Klein singularity of type $A$. This relation also implies that at infinite coupling $S$ is not a nilpotent operator. \\

\noindent For $N_f$ even, while $\mu_{N_f}^2$ is corrected in accordance to Eq. \ref{eq:generalcorrection} by appropriate contraction of S and M with gamma matrices, instead the product of S and M appearing in $\mu_{N_f-1}^2$ cannot be corrected by the spinor bilinear, and is thus constrained to be zero.

\item \textbf{Instantons}\\
Instantons by themselves can form meaningful constraints. These relations will have a nonzero charge under U(1) and will transform in some representation obtained by the symmetric product of the $I$ or $\tilde{I}$ representations. The symmetric product of the spinor representation has modularity 4:
\begin{equation}
    \text{Sym}^2\mu_{N_f}=\mu_{N_f}^2+\mu_{N_f-4}+\mu_{N_f-8}+\dots \ ,
\end{equation}
and similarly for the other spinor representation. We have found that the relation involving instantons constrain all the representation in the symmetric product to be zero, except the top one:
\begin{equation}
    \boxed{\text{Sym}^2I=(I\cdot I)|_{\mu_{N_f}^2}}
    \label{eq:instantonsymmetric}
\end{equation}
and similarly for $\tilde{I}$. The dot in Eq. \ref{eq:instantonsymmetric} represents some contraction with a gamma matrix. Such constraint is a Joseph-like relation; the Joseph relations determine closures of minimal nilpotent orbits and impose that all lower representation of the symmetric product of the adjoint representation (of any Lie algebra) should vanish, except the top component. The constraint in Eq. \ref{eq:instantonsymmetric} can be generalized by saying that the instanton of charge $k$ transforms in the top representation among the ones appearing in $\text{Sym}^k$. This follows from the computations of the quantum numbers of bare instantons, for the same theory, in \cite{hanany2025higgs}.

\end{itemize}

\noindent For the sake of clarity we group all constraints in the following table:
\renewcommand{\arraystretch}{1.3}
\begin{table}[H]
    \centering
    \begin{tabular}{|c|c|c|c|c|} \hline
      \textbf{Constraint}  & \textbf{R-charge} & \textbf{U(1) charge} & \textbf{Description} \\ \hline \hline
       $ M^2_{ij}=S^2\delta_{ij}$ & 4  & 0  & Casimir correction  \\ \hline 
      $(M\gamma_{})\cdot I=SI$ & $k+3$ & +1 & Eigenvalue equation \\ \hline
      $\epsilon\cdot \wedge^pM=S^a \, (\wedge^bM)$ & $2p$ & 0 & Hodge dual correction \\ \hline
      $ S^c \wedge ^dM_{ij}=(I\cdot\tilde{I)}|_{\mu_{2d}}$ & $2k+2$ & 0 & Instanton bilinear correction \\ \hline
      $\text{Sym}^2I=(I\cdot I)|_{\mu_{N_f}^2}$ & $2k+2$ & +2  & Instanton symmetric product \\ \hline
    \end{tabular}
    \caption{Chiral ring constraints for $Sp(k)$ with $0<N_f \leq 2k+3$ flavours in the vector representation, in the infinite coupling limit. The third constraint only appears for $N_f \leq 2k+1$; the parameters $a,b$ and $p$ are related by $N_f=p+b$ and $p=a+b$, with $p \leq k+1$. The parameters $c,d$ are related by $c+d=k+1$. The second and last constraint appear also for $\tilde{I}$, as explained previously, but are here omitted for readability.}
    \label{tab:infinitecouplingconstraintsGENERAL}
\end{table}

\noindent Let us point out that the pattern in Tab. \ref{tab:infinitecouplingconstraintsGENERAL} is quite natural, as it includes all that is allowed by $D_{N_f}$ and $U(1)$ symmetry. We should also note that for $Sp(k)$ with $N_f=2k+4$ and $N_f=2k+5$ there is symmetry enhancement and the constraints in table \ref{tab:infinitecouplingconstraintsGENERAL} acquire a different form, although the logic is the same. Their form will be explained in section \ref{sec:symmetryenhancement}.

\subsection*{Weak coupling}
As mentioned previously, at weak coupling the instantons are integrated out of the chiral ring. Therefore, they are put to zero in the constraints of Tab. \ref{tab:infinitecouplingconstraintsGENERAL}, which now have the following form:
\begin{table}[H]
    \centering
    \begin{tabular}{|c|c|c|} \hline
      \textbf{Constraint}  & \textbf{R-charge}  & $SO(2N_f)$ \textbf{rep.} \\ \hline \hline
       $ M^2_{ij}=S^2\delta_{ij}$ & 4   & $\mu_1^2+1$  \\ \hline 
      $\epsilon\cdot \wedge^pM=S^a \, (\wedge^bM)$ & $2p$  & $\mu_b$\\ \hline
      $ S^c \wedge ^dM_{ij}=0$ & $2k+2$  & $\mu_d$\\ \hline
    \end{tabular}
    \caption{Chiral ring constraints for $Sp(k)$ with $0<N_f \leq 2k+3$ flavours in the vector representation, at weak coupling. The parameters $a,b$ and $p$ are related by $N_f=p+b$ and $p=a+b$, with $p\leq k+1$, with the second constraint only appearing for $N_f\leq 2k+1$. Parameters $c,d$ satisfy $c+d=k+1$.}
    \label{tab:weakcouplingconstraintsGENERAL}
\end{table}
\noindent where we now specified the $SO(2N_f)$ representations, which we omitted in Tab. \ref{tab:infinitecouplingconstraintsGENERAL} not to clutter with the details depending on even/odd $N_f$ cases.\\

\noindent The constraints in Tab. \ref{tab:weakcouplingconstraintsGENERAL} are drastically different from the ones we would find using the F-term equations from the Lagrangian description of the theory. The weak coupling F-term equations for $Sp(k)$ with $N_f$ flavours \cite{ferlito2016tale} are:
 \begin{equation}
     \begin{array}{l}
          M^2=0  \\
          \text{rank}(M) \leq 2k
     \end{array}\label{eq:Fterms}
 \end{equation}
Indeed, assuming the Higgs branch at weak coupling to be the closure of a nilpotent orbit, all of the Casimir invariants of the meson matrix would be put to zero, which is clearly not the case in Tab. \ref{tab:weakcouplingconstraintsGENERAL}. Even at weak coupling, all the Casimir invariants are corrected by powers of $S$, as implied by the first row. The Pfaffian is also corrected by $S^{N_f}$, as can be seen by taking the second constraint and contracting both members by $b$ meson matrices.\\

\noindent All the relations in the second and third row of Tab. \ref{tab:weakcouplingconstraintsGENERAL} do not appear in the F-term equations, including the one obtained by putting $d=0$ in the last row, namely:
\begin{equation}
    S^{k+1}=0 \ .
\end{equation}
It is the nilpotency of this operator which is not taken into account by the F-terms, which put to zero all of its powers. The fermionic origin of the gaugino bilinear however does not imply $S$ should vanish altogether. The nilpotency of the gaugino bilinear implies that the holomorphic functions one can build on the Higgs branch at finite coupling can be split into disjoint sectors, namely as many as $h^{\vee}$. Each sector contains a given power of the gaugino bilinear. In the case of $k=1$, meaning $SU(2)$, which was studied in \cite{cremonesi2017instanton}, at weak coupling we would have:
\[
S^2=0
\]
hence there would be no difference between Tab. \ref{tab:weakcouplingconstraintsGENERAL} and Eqs. \ref{eq:Fterms}. This is likely the reason why the weak coupling corrections had not been visualized at the time. However, we conclude that the Higgs branches at finite coupling of Sp(k), with $k>2$, are not closures of nilpotent orbits of $D_{N_f}$.\\

\noindent The constraints in table \ref{tab:weakcouplingconstraintsGENERAL} also imply that the Higgs branch is perturbatively renormalized along the RG flow, despite widely held assumptions concerning non renormalization theorems. The theorems claim that, since the gauge coupling belongs to a background vector multiplet, it does not interfere with the Higgs branch. A loophole of the non renormalization theorem is that the Higgs branch is actually renormalized whenever massless degrees of freedom arise, such as in the infinite coupling limit. With the present work, we find that the chiral ring is affected all along the RG flow due to the gaugino bilinear.\\

\noindent It would be interesting to reproduce these constraints from a perturbative computation. In this perspective, we make an observation on the canonical vs supersymmetric normalization of the gauginos in the Lagrangian. The factor of $g^{-2}$ in the supersymmetric normalization can be absorbed in the definition of the gauginos, which then lead to:
\[
S \to g^4 S_{\text{can}}
\]
where $S_{\text{can}}$ is the gaugino bilinear built out of canonically normalized gauginos. Therefore, by substitution into Eq. \ref{tab:weakcouplingconstraintsGENERAL}, one could consider the first constraint as a 2-loop correction. The coefficients of the Casimir corrections are known to be non-zero, as computed in various cases in the next section.

\subsection*{Zero coupling}
At zero coupling the gaugino bilinear is removed from the chiral ring, whose relations are simply: 
\begin{table}[H]
    \centering
    \begin{tabular}{|c|c|c|} \hline
      \textbf{Constraint}  & \textbf{R-charge}  & $SO(2N_f)$ \textbf{rep.} \\ \hline \hline
       $ M^2_{ij}=0$ & 4    & $\mu_1^2+1$  \\ \hline 
      $  \wedge ^{k+1}M_{ij}=0$ & $2k+2$  & $\mu_d$\\ \hline
    \end{tabular}
    \caption{Chiral ring constraints for $Sp(k)$ with $0<N_f \leq 2k+3$ flavours in the vector representation, at zero coupling.}
    \label{tab:zerocouplingconstraintsGENERAL}
\end{table}
\noindent which agree with the F-term equations in Eq. \ref{eq:Fterms}.

\section{Case studies}\label{sec:methods}
We now explain the method used to derive the constraints presented above. The five dimensional theories we are studying live on a braneweb \cite{aharony1997branes,aharony1998webs} including $O7^-$ planes \cite{bergman20155d, bourget2020magnetic}. By tuning to zero all dimensionful parameters (whether dynamical or background fields), we are able to go to the origin of the moduli space of vacua. The Higgs branch at that point is then parameterized by transversal motion of consistent subwebs along the D7 branes \cite{cabrera2019tropical}. The structure of the Higgs branch, whether at the origin of the moduli space or at a generic point, is encoded into a quiver diagram, called a magnetic quiver.\\

\noindent The word magnetic is due to the fact that the massless degrees of freedom in the 5d theory along the Higgs branch are excitations of virtual D3 branes stretched between the consistent subwebs. As all codimension 3 objects, the D3 branes are 't Hooft Polyakov monopoles from the point of view of the 5d theory. We can thus evaluate the chiral ring of the moduli space of the magnetic monopole operators\footnote{Strictly speaking, there is a logical leap from considering 't Hooft Polyakov monopoles, which are classical solutions, to magnetic monopole operators, which are local disoder operators in the path integral. A proof justifying microscopically this procedure is still lacking. However, this technique has produced results which agree with alternative methods, for example \cite{bourget2020brane}.} using the monopole formula \cite{cremonesi2014monopole}, which by construction corresponds to the chiral ring of the Higgs branch of the theory at hand.\\

\noindent Such procedure allows us to find the Hilbert series (HS) of the chiral ring \cite{benvenuti2007counting}, meaning the generating function of 1/2 BPS local gauge invariant operators, graded by their R-charge. From this series, we can find the independent generators and relations by using the plethystic logarithm (PL) \cite{feng2007counting, gray2008sqcd}. In particular, the generators correspond to the first few positive terms of the PL and the constraints correspond to the first few negative terms. If the space is not what mathematicians call a “complete intersection", additional relations will appear among the relations themselves, called syzygies, which makes much more difficult the search for the independent set of relations. One way to ascertain whether the found set of constraints is complete is to use tools such as Macaulay2, which allows to compute the HS given the generators and their constraints (which span the ideal of the ring). In practice, the use of Macaulay2 is limited by the number of generators it can handle, so while useful as a consistency check in selected cases, it does not provide a fully systematic verification.\\

\noindent We did not use the superconformal index to compute any of the generators or constraints as, at least in 5d, the HS has proved to be a more streamlined tool to study the moduli space of vacua. Previous computations involving the index, among which \cite{kim20125, bergman2014discrete}, could only find the first few orders in its perturbative expansion, dealing with some ambiguities in the Nekrasov instanton partition function, and could only tell the dimension and flavour symmetry of the Higgs branch. We have found the HS to be more informative, in this context, as it provides a direct and unambiguous handle on the whole chiral ring, making it the more powerful tool for our purposes.\\

\noindent Let us point out that the above-described technique is only sensible to the continuous part of the moduli space of vacua, and in principle does not account for the discrete part. For example, if we compute the chiral ring at finite coupling using such procedure, we will only find the mesons, with some constraint. We would not find the gaugino bilinear and the full set of relations. Therefore, in order to find the full information on the Higgs branch, it is more informative to compute the infinite coupling chiral ring and then flow along the RG flow by decoupling the instantons. In summary, the following two procedures do not commute:
\begin{table}[H]
    \centering
    \begin{tabular}{|c|c|c|}
        Braneweb at infinite coupling & & Braneweb at infinite coupling \\
        $\downarrow$ &  &   $\downarrow$\\
        Deformation to finite coupling & $ \qquad \neq \qquad $ & Hilbert series extraction\\
        $\downarrow$ & &   $\downarrow$\\
        Hilbert series extraction & & Flow to finite coupling\\
    \end{tabular}
    \caption{Non commutativity of the two procedures to extract the finite coupling chiral ring. Only the procedure on the right yields the complete chiral ring and relations.}
    \label{tab:proceduresdonotcommute}
\end{table}

\noindent We will now examine a few instructive cases for $Sp(2)$ in order to understand how to apply the infinite coupling constraints in Tab. \ref{tab:infinitecouplingconstraintsGENERAL}. We will pick some amount of flavours where all kind of relations in Tab. \ref{tab:infinitecouplingconstraintsGENERAL} are displayed. We will also treat the low number of flavours, where interesting features take place; at the end we are going to examine the flavours that allow symmetry enhancement.\\

\noindent A few words about notation. In the following examples, mesons always transform in the adjoint representations of $SO(2N_f)$ hence they will be denoted by antisymmetric matrices $M_{[ij]}$. Instantons of U(1) charge $\pm 1$ transform in the spinor representation, which is complex for odd $N_f$ and real or presudoreal for even $N_f$, and is denoted by letters in the greek alphabet. In particular, for real or pseudoreal representation, spinor indices exist in two sets, dotted and undotted. The representations of $SO(2N_f)$ will always be denoted by their highest weights, as before, rather than with Dynkin labels. Furthermore, in the quiver diagrams, a dotted line represent a leg of gauge nodes whose rank increase monotonically by one unit to match the endpoints of the leg.

\subsection{$Sp(2) \text{ with } N_f=5$}
\paragraph{Finite coupling: incomplete technique}
We will now compute the chiral ring at finite coupling using the “incomplete" technique, namely the one on the left of Tab. \ref{tab:proceduresdonotcommute}. We will then compute the infinite coupling chiral ring, flow to finite coupling and show the full set of generators and relations.\\

\noindent The Higgs branch of Sp(2) with $N_f=5$ is $\bar{\mathcal{O}}_D^{[2^4,1^2]}$, represented by the quiver:
\begin{figure}[H]
    \centering
  \begin{tikzpicture}

  \filldraw (0,0) circle (2pt) node[below=2pt] {1};
  \filldraw (2,0) circle (2pt) node[right=2pt and 2pt] {3};

  \draw[dotted, thick] (0,0) -- (2,0);

  \filldraw (3.3,1.2) circle (2pt) node[above=4pt] {2};
  \filldraw (3.3,-1.2) circle (2pt) node[below=4pt] {2};

  \draw (2,0) -- (3.3,1.2);
  \draw (2,0) -- (3.3,-1.2);

  \filldraw (4.5,0) circle (2pt) node[right=2pt] {1};

  \draw (3.3,1.2) -- (4.5,0);
  \draw (3.3,-1.2) -- (4.5,0);

\end{tikzpicture}
    \caption{Higgs branch of Sp(2) at finite coupling with $N_f=5$.}
    \label{fig:finitecouplinNf=5}
\end{figure}
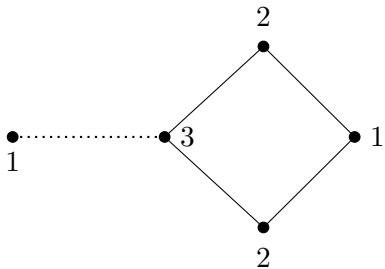
\noindent The global symmetry of the theory is $SO(10)\times U(1)$, while the global symmetry of the quiver in Fig. \ref{fig:finitecouplinNf=5} is only $SO(10)$; indeed the technique described in Tab. \ref{sec:methods} is intrinsically not sensible to discrete vacua in the theory, hence it does not detect the gaugino bilinear, as explained earlier. The Hilbert series is:
\begin{equation}
    \begin{array}{ll}
       \text{HS}(t)=  & 1+\mu_2t^2+ \\
         & +(\mu_2^2+\mu_4\mu_5)t^4+\dots
    \end{array}
\end{equation}
whose PL is:
\begin{equation}
    \begin{array}{ll}
        \text{PL[HS]}(t) &  = \mu_2 t^2+\\
         & -(1+\mu_1^2)t^4+\dots
    \end{array}
\end{equation}
hence the unique generator appears to be:
\begin{table}[H]
    \centering
    \begin{tabular}{|c|c|c|c|} \hline
       Generators  &  R charge & $SO(10)$ & U(1) \\ \hline
        $M_{[ij]}$ & 2 & $\mu_2$ & 0 \\  \hline
    \end{tabular}
    \caption{Finite coupling generators for $N_f=5$ using the incomplete method in \ref{tab:proceduresdonotcommute}. The gaugino bilinear is indeed missing, as expected.}
    \label{tab:finite coupling generators for Nf=5}
\end{table}
\noindent The constraint appears to be:
\begin{table}[H]
    \centering
    \begin{tabular}{|c|c|c|c|} \hline
       Constraints  &  R charge & $SO(10)$ & U(1) \\ \hline
        $M^2_{ij}=0$ & 4 & $\mu_1^2+1$ & 0 \\ \hline
    \end{tabular}
    \caption{Incomplete set of finite coupling constraints for $N_f=5$.}
    \label{tab:finite coupling constraints for Nf=5}
\end{table}
\noindent which is indeed incomplete, as we are now going to see.
\paragraph{Infinite coupling}
\noindent The Higgs branch at infinite coupling has global symmetry $SO(10)\times U(1)$ and is represented by the quiver:
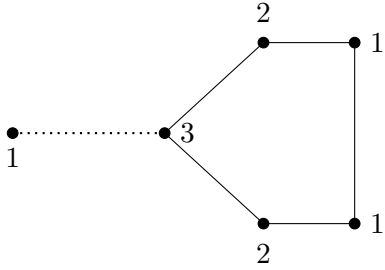
\begin{figure}[H]
    \centering
  \begin{tikzpicture}

  \filldraw (0,0) circle (2pt) node[below=2pt] {1};
  \filldraw (2,0) circle (2pt) node[right=2pt and 2pt] {3};

  \draw[dotted, thick] (0,0) -- (2,0);

  \filldraw (3.3,1.2) circle (2pt) node[above=4pt] {2};
  \filldraw (3.3,-1.2) circle (2pt) node[below=4pt] {2};

  \draw (2,0) -- (3.3,1.2);
  \draw (2,0) -- (3.3,-1.2);

  \filldraw (4.5,1.2) circle (2pt) node[right=2pt] {1};

  \filldraw (4.5,-1.2) circle (2pt) node[right=2pt] {1};

  \draw (3.3,1.2) -- (4.5,1.2);
  \draw (3.3,-1.2) -- (4.5,-1.2);
  \draw (4.5,1.2) -- (4.5,-1.2);

\end{tikzpicture}
    \caption{Higgs branch of Sp(2) at infinite coupling with $N_f=5$}
    \label{fig:infinitecouplinNf=5}
\end{figure}
\noindent which belongs to $E_4$ exceptional sequence \cite{ferlito20183d}. The change from finite coupling to infinite coupling can be clearly seen in the Hasse diagram, where the top slice $d_3$ is enhanced to a $a_4$:\\
\begin{figure}[H]
    \centering
   \begin{tikzpicture}[every node/.style={font=\small}]

  \filldraw (0,2) circle (2pt); 
  \filldraw (0,1) circle (2pt); 
  \filldraw (0,0) circle (2pt); 
  \draw (0,0) -- (0,2);
  \node[left=4pt] at (0,1.5) {$d_3$}; 
  \node[left=4pt] at (0,0.5) {$d_5$}; 

  \filldraw (2,2) circle (2pt); 
  \filldraw (2,1) circle (2pt); 
  \filldraw (2,0) circle (2pt); 
  \draw (2,0) -- (2,2);
  \node[right=4pt] at (2,1.5) {$a_4$}; 
  \node[right=4pt] at (2,0.5) {$d_5$}; 

\end{tikzpicture}
    \caption{Finite coupling (left) and infinite coupling (right) Hasse diagrams for Sp(2) with 5 flavours.}
    \label{fig:HasseNf=5}
\end{figure}
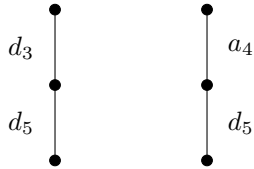

\noindent We recall that the spinor representation of SO(10) is complex and can be contracted with:
\[
\delta_{\alpha}^{\beta} \quad (\gamma^i)_{\alpha\beta} \quad (\gamma^{ij})_{\alpha}^{\beta} \quad \text{etc.}
\]
The Hilbert series up to sixth order is:
\begin{equation}
    \begin{array}{ll}
       \text{HS}(t)  & =1+(1+\mu_2)t^2+ \\
         & +(q\mu_5+q^{-1}\mu_4)t^3+\\
         & +(1+\mu_2+\mu_2^2+\mu_4\mu_5)t^4+\\
         & +(q(\mu_5+\mu_2\mu_5)+q^{-1}(\mu_4+\mu_2\mu_4))t^5+\\
         & +(1+\mu_1+\mu_2+\mu_4\mu_5\dots)t^6+\dots
    \end{array}
\end{equation}
and its PL is:
\begin{equation}
    \begin{array}{ll}
        \text{PL[HS]}(t) & = (1+\mu_2)t^2+\\
         & +(q\mu_5+q^{-1}\mu_4)t^3+ \\
         & -(1+\mu_1^2)t^4+\\
         & -(q(\mu_5+\mu_1\mu_4)+q^{-1}(\mu_4+\mu_1\mu_5))t^5+\\
         & -(1+\mu_2+2\mu_4\mu_5+(q^2+q^{-2})\mu_1)t^6+\dots
    \end{array}
\end{equation}
The full set of generators is:
\begin{table}[H]
    \centering
    \begin{tabular}{|c|c|c|c|} \hline
       Generators  &  R charge & $SO(10)$ & U(1) \\ \hline
        $M_{[ij]}$ & 2 & $\mu_2$ & 0 \\ 
        $S$ & 2 & $1$ & 0 \\
        \hline
        $I_{\alpha}$  & 3 & $\mu_5$ & $1$\\
        $\tilde{I}^{\alpha}$ & 3 & $\mu_4 $&$-1$ \\ \hline
    \end{tabular}
    \caption{Infinite coupling generators for $N_f=5$}
    \label{tab:infinite coupling generators for Nf=5}
\end{table}
\noindent with the following constraints:
\begin{table}[H]
    \centering
    \begin{tabular}{|c|c|c|c|} \hline
       Constraints  &  R charge & SO(10) & U(1) \\ \hline
        $M^2_{ij}= S^2\delta_{ij}/10$ & 4 & $\mu_1^2+1$ & 0 \\ 
      \hline
       $ S I_{\beta}(\gamma_i)^{\beta\alpha}+ M_{[ij]}I_{\beta}(\gamma_{j})^{\alpha \beta}=0$ & 5 & $\mu_1\mu_4+\mu_5$ & $1$
       \\  
       $ S \tilde{I}^{\alpha}(\gamma_i)_{\beta\alpha}+ M_{[ij]}\tilde{I}^{\alpha}(\gamma_{j})_{\alpha \beta}=0$ & 5 & $\mu_1\mu_5+\mu_4$ & $-1$
       \\  \hline
       $S^3+I_{\alpha}\delta^{\alpha}_{\beta}\tilde{I}^{\beta}=0$ & 6 & $1$ & 0 \\
$S^2M_{ij}+(\gamma_{ij})^{\alpha}_{\beta}I_{\alpha}\tilde{I}^{\beta}=0$ & 6 &$ \mu_2$ & 0 \\
     $ SM_{[ij}M_{kl]}+I_{\alpha}\tilde{I}^{\beta}(\gamma_{ijkl})_{\alpha}^{\beta}=0$& 6 & $\mu_4\mu_5$ & 0 \\ 
     
 $M_{[ij}M_{kl}M_{mn]}\epsilon_{ijklmnspqr}+ I_{\alpha}\tilde{I}^{\beta}(\gamma_{spqr})_{\beta}^{\alpha}=0$& 6 & $\mu_4\mu_5$ & 0 \\
    $I_{\alpha}I_{\beta}(\gamma_i)^{\alpha\beta}=0$ & 6  & $\mu_1$ & 2 \\
  $\tilde{I}^{\alpha}\tilde{I}^{\beta}(\gamma_i)_{\alpha\beta}=0$ & 6  & $\mu_1$ & $-2$ \\

 \hline
    \end{tabular}
    \caption{Infinite coupling constraints for $N_f=5$}
    \label{tab:infinite coupling constraints for Nf=5}
\end{table}
\noindent We can see in this table all the type of constraints of Tab. \ref{tab:infinitecouplingconstraintsGENERAL}. There is the familiar trace correction and the eigenvalue equations. Furthermore, we find the correction to $S (\wedge M)$ by the instanton bilinear, which in this case transforms in the representations:
\[
\mu_5\otimes \mu_4=\mu_4\mu_5+\mu_2+1 \ .
\]
We also have the Hodge dual constraints, as there are two constraints in the $\mu_4\mu_5$ involving three different operators. Finally, we find the condition on the symmetric product of the instantons, which now resembles a pure spinor condition. We can obtain the constraints at finite coupling by putting to zero the instantons, thus noticing that the set in Tab. \ref{tab:finite coupling constraints for Nf=5} is incomplete; in turn, we can recover the constraints at zero coupling by putting to zero the gaugino bilinear.\\

\noindent The generalization to higher rank is relatively easy. For general $Sp(k)$, the instantons have R-charge equal to the dual Coxeter number $h^{\vee}=k+1$, hence the relations which previously appeared at order 5 will now appear at order $k+3$. Furthermore, the relations that previously appeared at order 6, namely the instanton bilinear relations, will appear at order $2k+2$.
\subsection{$Sp(2)$ with $N_f=4$}
The Higgs branch of the theory at infinite coupling has global symmetry $SO(8)\times U(1)$, it is the union of two cones, with a non trivial intersection:
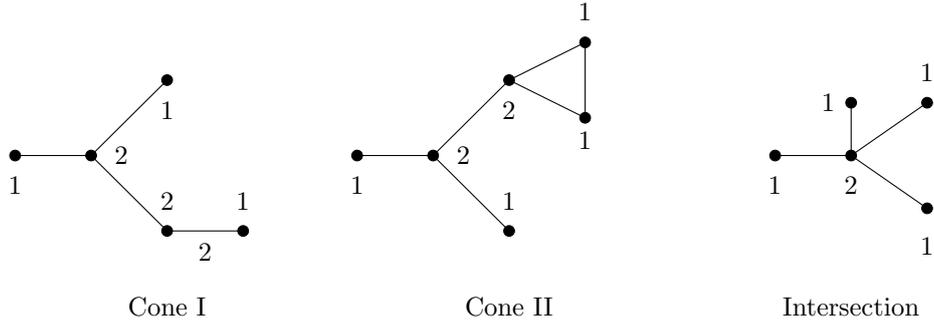
\begin{figure}[H]
    \centering
  \begin{tikzpicture}[scale=1, every node/.style={font=\small}]

  \begin{scope}[shift={(-3,2)}]
  \filldraw (3,1) circle (2pt);                  
  \filldraw (2,0) circle (2pt);                  

  \node at (3,0.6) {1};
  \node at (2.4,0) {$2$};

  \draw (3,1) -- (2,0);

  \filldraw (1,0) circle (2pt);                
  \filldraw (3,-1) circle (2pt);               

  \node at (1,-0.4) {1};
  \node at (3,-0.6) {2};

  \draw (1,0) -- (2,0);
  \draw (2,0) -- (3,-1);

  \filldraw (4,-1) circle (2pt);
  \node at (4,-0.6) {1};
  \draw (3,-1) -- node[midway, below=1pt] {2} (4,-1);

  \node at (3,-2) {Cone I};
  \end{scope}
  
 \begin{scope}[shift={(3.5,2)}]
  \filldraw (1,1) circle (2pt);                  
  \filldraw (0,0) circle (2pt);                  

  \node at (1,0.6) {2};
  \node at (0.4,0) {$2$};

  \draw (1,1) -- (0,0);

  \filldraw (-1,0) circle (2pt);                
  \filldraw (1,-1) circle (2pt);               

  \node at (-1,-0.4) {1};
  \node at (1,-0.6) {1};

  \draw (-1,0) -- (0,0);
  \draw (0,0) -- (1,-1);

  \filldraw (2,1.5) circle (2pt);
  \node at (2, 1.9) {1};
  \draw (1,1) -- node[midway, below=1pt] {} (2,1.5);
    \filldraw (2,0.5) circle (2pt);
  \node at (2, 0.2) {1};
  \draw (1,1) -- node[midway, below=1pt] {} (2,0.5);
  \draw (2,1.5) -- (2,0.5);
    \node at (1,-2) {Cone II};
  \end{scope}

   \begin{scope}[shift={(7,2)}]
  \filldraw (1,0) circle (2pt);                  
  \filldraw (2,0) circle (2pt);                  

  \node at (1,-0.4) {1};
  \node at (2,-0.4) {2};

  \draw (1,0) -- (2,0);

  \filldraw (3,0.7) circle (2pt);                
  \filldraw (3,-0.7) circle (2pt);               

  \node at (3,1.1) {1};
  \node at (3,-1.2) {1};

  \draw (2,0) -- (3,0.7);
  \draw (2,0) -- (3,-0.7);

  \filldraw (2,0.7) circle (2pt);
  \node at (1.7,0.7) {1};
  \draw (2,0) -- (2,0.7);

    \node at (2,-2) {Intersection};
  \end{scope}
  
\end{tikzpicture}
    \caption{Infinite coupling Higgs branch of Sp(2) with $N_f=4$}
    \label{fig:quiverinfinitecouplingNf=4}
\end{figure}
\noindent where the first cone is unchanged with respect to the Higgs branch at finite coupling, while the second one is enhanced. Such enhancement can be clearly seen from the Hasse diagrams:
\begin{figure}[H]
    \centering
   \begin{tikzpicture}[every node/.style={font=\small}]

  \filldraw (-0.5,2) circle (2pt); 
    \filldraw (0.5,2) circle (2pt); 
  \filldraw (0,1) circle (2pt); 
  \filldraw (0,0) circle (2pt); 
  \draw (0,0) -- (0,1) -- (0.5,2);
  \draw (0,1) -- (-0.5,2);
  \node[left=4pt] at (-0.1,1.5) {$a_1$}; 
  \node[left=4pt] at (1,1.5) {$a_1$}; 
  \node[left=4pt] at (0.1,0.5) {$d_4$}; 

  \filldraw (1.5,2) circle (2pt); 
    \filldraw (2.5,2) circle (2pt); 
  \filldraw (2,1) circle (2pt); 
  \filldraw (2,0) circle (2pt); 
  \draw (2,0) -- (2,1) -- (2.5,2);
  \draw (2,1) -- (1.5,2);
  \node[left=4pt] at (1.9,1.5) {$a_1$}; 
  \node[left=4pt] at (3,1.5) {$a_2$}; 
  \node[left=4pt] at (2.1,0.5) {$d_4$}; 

\end{tikzpicture}
    \caption{Finite coupling (left) and infinite coupling (right) Hasse diagrams for Sp(2) with 4 flavours. The bifurcation is due to the existence of two cones.}
    \label{fig:HasseNf=5}
\end{figure}
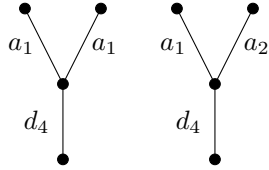

\noindent We recall that the spinor representation of SO(8) is real and can be contracted, for example, by:
\[
\delta_{\alpha\beta}\quad \delta_{\dot{\alpha}\dot{\beta}} \quad (\gamma^i)_{\alpha \dot{\alpha}} \quad (\gamma^{ij})_{\alpha\beta} \quad \text{etc.}
\]
\noindent The Hilbert series up to the sixth order is:
\begin{equation}
    \begin{array}{ll}
        \text{HS}(t) & = 1+\mu_2 t^2+  \\
         & + (q+q^{-1})\mu_4t^3+\\
         & + (1+\mu_2+\mu_2^2+\mu_4^2+\mu_3^2)t^4+\\
         & +(q+q^{-1})(\mu_4+\mu_2\mu_4)t^5+\\
         & +(1+\mu_2+(1+q^2+q^{-2})\mu_4^2+\dots )t^6+ \dots
    \end{array}
\end{equation}
with the PL being:
\begin{equation}
    \begin{array}{ll}
        \text{PL[HS]}(t) & =  1+\mu_2t^2+  \\
         &  + (q+q^{-1})\mu_4t^3+\\
         & -(1+\mu_1^2)t^4+ \\
         & -(q+q^{-1})(\mu_4+\mu_1\mu_3)t^5 \\
         & -(1+q^2+q^{-2}+2\mu_2+\mu_4^2+\mu_3^2)t^6+(\dots)t^6+ \dots
    \end{array}
\end{equation}
by which we conclude that the generators are:
\begin{table}[H]
    \centering
    \begin{tabular}{|c|c|c|c|}\hline
       Generators  &  R-charge & $SO(8)$ & U(1) \\ \hline
      $M_{[ij]}$ & 2 & $\mu_2$ & 0 \\ 
       S & 2 & 1 & 0 \\ \hline
       $I_{\alpha}$ & 3 & $\mu_4$ & 1 \\
       $\tilde{I}_{\alpha}$ & 3 & $\mu_4$ & -1 \\ \hline
    \end{tabular}
    \caption{Infinite coupling generators for $N_f=4$}
\label{tab:infinitecouplinggeneratorsforNf=4}
\end{table}
\noindent and the constraints are:
\begin{table}[H]
    \centering
    \begin{tabular}{|c|c|c|c|}\hline
      Constraint  &  R-charge & SO(8) & U(1) \\ \hline
      $M^2_{ij}=S^2\delta_{ij}/8$ & 4 & $\mu_1^2+1$ & 0 \\ \hline
      $SI_{\alpha}(\gamma_i)_{\alpha \dot{\beta}}+M_{ij}I_{\alpha}(\gamma_j)_{\alpha\dot{\beta}}=0$ & 5 & $\mu_1\mu_3+\mu_4$ & $+1$ \\
      $S\tilde{I}_{\alpha}(\gamma_i)_{\alpha \dot{\beta}}+M_{ij}\tilde{I}_{\alpha}(\gamma_j)_{\alpha\dot{\beta}}=0$ & 5 & $\mu_1\mu_3+\mu_4$ & $-1$ \\ \hline
       $S^3+ I_{\alpha}\tilde{I}_{\beta} \delta^{\alpha\beta}=0$ & 6 & $1$ & 0 \\
         $S^2M_{ij}+I_{\alpha}\tilde{I}_{\beta}(\gamma_{ij})^{\alpha\beta}=0$ & 6 & $\mu_2$ & 0 \\
           $S^2M_{rt}+M_{[ij}M_{mn}M_{kl]}\epsilon_{ijmnklrt}=0$ & 6 & $\mu_2$ & 0 \\
         $SM_{[ij}M_{kl]}(\gamma^{ijkl})_{\alpha\beta}+(I_{\alpha}\tilde{I}_{\beta}-S^3\delta_{\alpha\beta}/8)=0$ & 6 & $\mu_4^2$ & 0 \\
       $SM_{[ij}M_{kl]}(\gamma^{ijkl})_{\dot{\alpha}\dot{\beta}}=0$   & 6 & $\mu_3^2$ & 0 \\
        $I^2=0$ & 6 & $1$ &  2\\
         $\tilde{I}^2=0$ & 6 & $1$ &  $-2$\\
        \hline
    \end{tabular}
    \caption{Constraints at infinite coupling for $N_f=4$}
    \label{tab:infinitecouplingconstraintsNf=4}
\end{table}
\noindent As usual we find the trace correction and the eigenvalue equation. We also find the correction of $S(\wedge M)$ by the instanton bilinears and, being in the case $N_f\leq 2k+1$, we also have an instance of Hodge dual correction, as there are three operators trasforming in the $\mu_2$ representation. As usual with even colours, we also have the condition on the opposite type of spinor, namely the condition in the representation $\mu_3^2$. At the end we also account for the two nilpotency condition on the symmetric products of $I$ and $\tilde{I}$. 

\paragraph{Fixing the coefficients}
In Tab. \ref{tab:infinitecouplingconstraintsNf=4}, the PL only fixes the representations in which the constraint appear, however the coefficients of the operators in the relations might be, a priori, vanishing. To give an example, the PL only tells us that there is one singlet operator at order 4 in the HS, and one singlet constraint at order 4 in the PL. Being however possible to write two such singlets at order 4, they have to be constrained as:
\[
a \cdot \text{tr}M^2+b \cdot S^2=0 \ .
\]
This does not exclude that either $a$ or $b$ could be zero. For $N_f=4$ and lower flavours we are able to fix the coefficients and to show that they are finite. For higher number of flavour we are unaware of any construction allowing to show such a thing, however, for consistency with the low flavour cases, we argue for them to be finite as well.\\

\noindent The way we fix these coefficients is the following, using techniques developed in \cite{hanany1999issues}. We consider cone II of Fig. \ref{fig:quiverinfinitecouplingNf=4} as an abstract moduli space, without relation to its 5d origin. Such moduli space also happens to be the Coulomb branch of a 3d $\mathcal{N}=4$ theory described by the quiver itself. By 3d mirror symmetry, we can find the quiver theory whose Higgs branch is isomorphic to cone II. Such quiver theory is the following: 

\begin{figure}[H]
    \centering
 \begin{tikzpicture}[every node/.style={font=\small}]
  \filldraw (0,0) circle (2pt) node[below=2pt] {Sp(1)}; 
  \filldraw (2,0) circle (2pt) node[below=2pt] {U(2)}; 

  \draw (0,0) -- (2,0);

  \node[draw, minimum size=4mm] (topL) at (0,1.2) {};
  \node[above=2pt of topL] {4};

  \node[draw, minimum size=4mm] (topR) at (2,1.2) {};
  \node[above=2pt of topR] {1};

  \draw (0,0) -- (topL);
  \draw (2,0) -- (topR);
\end{tikzpicture}
    \caption{3d mirror theory of enhanced cone in Fig. \ref{fig:quiverinfinitecouplingNf=4}, seen as a 3d $\mathcal{N}=4$ theory.}
    \label{fig:mirrorNf=4}
\end{figure}
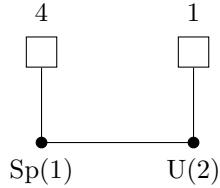
\noindent Let us label\footnote{The SO(8) representation is denoted by the highest weight fugacity, as previously, while the other representation are denoted by Dynkin labels. The representation $1$ of SO(8) is the singlet, which is different from, say, $[1]$ of Sp(1), which is a doublet.} the fields and their transformation laws under the symmetries:
\begin{table}[H]
    \centering
    \begin{tabular}{|c|c|c|c|c|} \hline
       Matter content & $Sp(1)$ & $U(2)$ & $SO(8)$ & $U(1)$ \\ \hline
       $A^{\alpha}_i$  & $[1]$ & $[0]_0$ & $\mu_1$ &  $0$ \\ \hline
       $B^{\alpha}_a$ & $[1]$ & $[1]_1$ & $1$ & 0 \\
       $\tilde{B}^{\beta,b}$ & $[1]$ & $[1]_{-1}$ & $1$ & 0 \\ \hline
       $C^{a}$ & $[0]$ & $[1]_{-1}$ & $1$ & $+1$ \\
       $\tilde{C}_{b}$ & $[0]$ & $[1]_{1}$ & $1$ & $-1$ \\ \hline
       $(\phi_1)_{(\alpha \beta)}$ & $[2]$ & $[0]_0$ & $1$ & 0 \\
       $(\phi_2)_{a}^{b}$ & $[0]$ & $[2]_0$ & $1$ & 0 \\ \hline
    \end{tabular}
    \caption{Quantum numbers of matter fields of the theory in Fig. \ref{fig:mirrorNf=4}.}
    \label{tab:mattercontent}
\end{table}
\noindent We can thus write the following gauge invariant combinations, where we identify the generators of Tab. \ref{tab:infinitecouplinggeneratorsforNf=4}: 
\begin{table}[H]
    \centering
    \begin{tabular}{|c|c|} \hline
        Generator & Definition \\ \hline
        $M_{[ij]}$ & $A_i^{\alpha} A_j^{\beta} \Omega_{\alpha\beta}$ \\
        $S$ & $\tilde{C}_{b}C^{a}\delta_{a}^{b}$  \\
        $I^i$ & $\Omega_{\alpha\beta}A^{i,\alpha}B^{\beta}_a C^{a}$ \\
        $\tilde{I}^j$ & $\Omega_{\alpha\beta}A^{j,\alpha}\tilde{B}^{\beta,b}\tilde{C}_{b} $\\ \hline
    \end{tabular}
    \caption{Generators of the Higgs branch of the quiver gauge theory in Fig. \ref{fig:mirrorNf=4}, also corresponding to the generator of the Higgs branch of the quiver in Fig. \ref{fig:quiverinfinitecouplingNf=4}.}
    \label{tab:generatorsalternativeset}
\end{table}
\noindent We can write the following superpotential, which is gauge and flavour invariant:
\begin{equation}
    W= A^{i,\alpha} A^{j,\beta}(\phi_1)_{\alpha\beta}\delta_{ij}+(\phi_1)_{\alpha\beta}B^{\alpha}_a \tilde{B}^{b,\beta}\delta^{a}_{b}+(\phi_2)_{b}^{a}B^{\alpha}_a \tilde{B}^{b,\beta}\Omega_{\alpha\beta}+(\phi_2)_{a}^{b}\tilde{C}_{b}C^{a} \ ,
\end{equation}
which gives rise to the following F-term equations:
\begin{equation}
    \begin{array}{ll}
        \dfrac{\partial W}{\partial \phi_1}= & A^{i\alpha}A^{j \beta}\delta_{ij}+\dfrac{1}{2}(B^{\alpha}_b\tilde{B}^{a,\beta}\delta^b_a+B^{\beta}_b\tilde{B}^{a,\alpha}\delta^b_a)=0 \\
        & \\
        \dfrac{\partial W}{\partial \phi_2} =  & B^{\alpha}_a \tilde{B}^{b,\beta}\Omega_{\alpha\beta}+\tilde{C}_{a}C^{b}=0
    \end{array}
\end{equation}
Using these equations it is easy to show that:
\[
M_{ij}M_{ij}=A_i^{\alpha}A_j^{\beta}\Omega_{\alpha\beta}A_i^{\gamma}A_j^{\delta} \Omega_{\gamma \delta}= S^2
\]
implying that the coefficients are indeed non vanishing. In a similar way, one can prove the other constraints in Tab. \ref{tab:infinitecouplingconstraintsNf=4}.

\subsection{Low number of flavours}
We will now examine the instructive cases with lower number of flavours.

\subsection*{$N_f=2$}
In this case it is interesting to see the growth of the Higgs branch from finite coupling to infinite coupling.
\paragraph{Finite coupling: incomplete technique} The Higgs branch at finite coupling is the union of two cones which intersect at the origin, both of which are $\mathbb{C}^2/\mathbb{Z}_2$, i.e. $\mathbf{A}_1$ singularities. The magnetic quiver of each cone is:
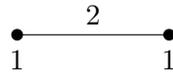
\begin{figure}[H]
    \centering
    \begin{tikzpicture}

  \filldraw (0,0) circle (2pt) node[below=2pt] {1};
  \filldraw (2,0) circle (2pt) node[below=2pt] {1};

  \draw (0,0) -- (2,0) node[midway, above] {2};

\end{tikzpicture}
    \caption{Single cone of Higgs branch of Sp(2) at infinite coupling with $N_f=2$}
    \label{fig:enter-label}
\end{figure}
\noindent The global symmetry is thus $SU(2)\times SU(2)\times U(1)$, but the $U(1)$ does not appear at finite coupling due to limitations of the technique. We will consider $SO(4)$, instead of $SU(2)\times SU(2)$, out of coherence with the higher flavour cases. Notice that the spinor representation of SO(4) is pseudoreal and its indices $ \dot{\alpha},\alpha$ can be raised or lowered with an epsilon tensor or with the gamma matrices:
\[
(\gamma^a)_{\alpha\dot{\alpha}} \qquad (\gamma_{ab})_{\alpha \beta} \quad \text{etc.}
\]
\noindent The Hilbert series up to fourth order is:
\begin{equation}
    \text{HS}(t)=1+(\mu_1^2+\mu_2^2)t^2+(\mu_1^4+\mu_2^4)t^4+\dots \ .
\end{equation}
The PL is:
\begin{equation}
    \text{PL}[\text{HS}]=(\mu_1^2+\mu_2^2)t^2-(2+\mu_1^2\mu_2^2)t^4+\dots
\end{equation}
The generators of the continuous part of the Higgs branch are the self-dual and anti self-dual components of the adjoint $M_{[ij]}$ of SO(4), which are 4 by 4 matrices with 3 real entries each, defined as:
\begin{equation}
\begin{array}{cccc}
   M_{ij}^+  & =M_{[ij]}+\epsilon_{ijkl}M^{kl}/2 & \quad \text{with}  & \quad M_{ij}^+=(\gamma_{ij})^{\alpha\beta}M_{\alpha\beta}^+ \\
   M_{ij}^-  & =M_{[ij]}-\epsilon_{ijkl}M^{kl}/2 & \quad \text{with} &  \quad M_{ij}^-=(\gamma_{ij})^{\dot{\alpha}\dot{\beta}}M_{\dot{\alpha}\dot{\beta}}^- \\
\end{array}\label{eq:mplusmmins}
\end{equation}
such that:
\[
M_{ij}=(M_{ij}^+
+M_{ij}^-)/2 \ .\]
The generators from the incomplete technique are thus: 
\begin{table}[H]
    \centering
    \begin{tabular}{|c|c|c|c|} \hline
       Generators  & R charge & $SO(4)$ & U(1) \\ \hline
        $M_{\alpha\beta}^+$ & 2 & $\mu_2^2$ & 0 \\
        $M_{\dot{\alpha}\dot{\beta}}^-$ & 2 & $\mu_1^2$ & 0 \\ \hline
    \end{tabular}
    \caption{Incomplete set of finite coupling generators for $N_f=2$}
    \label{tab:finitecouplingNf=2}
\end{table}
\noindent and the three constraints appear to be:
\begin{table}[H]
    \centering
    \begin{tabular}{|c|c|c|c|} \hline
      Constraints  & R charge & $SO(4)$ & U(1) \\ \hline
    $\text{tr}M_+^2=0$ & 4 & $1$ & 0 \\
    $\text{tr}M_-^2=0$ & 4 & $1$ & 0 \\ 
    $M_{\alpha\beta}^+M_{\dot{\gamma}\dot{\delta}}^-=0$ & 4 & $\mu_1^2\mu_2^2$ & 0 \\ \hline
    \end{tabular}
    \caption{Incomplete set of finite coupling constraints for $N_f=2$}
    \label{tab:constraintsNf=2finitecoupling}
\end{table}
\noindent signaling that we have two orthogonal $\mathbf{A}_1$ singularities that meet at the origin. Again, this procedure is incomplete as it does not account for the contribution of $S$. 

\paragraph{Infinite coupling}
The Higgs branch of theory at infinite coupling is still the union of two cones, one of which is exactly the same as the classical case, i.e. an $\mathbf{A}_1$ singularity, while the other one is represented by:
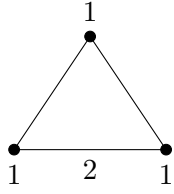
\begin{figure}[H]
    \centering
  \begin{tikzpicture}

  \filldraw (0,0) circle (2pt) node[below=2pt] {1};       
  \filldraw (2,0) circle (2pt) node[below=2pt] {1};       
  \filldraw (1,1.5) circle (2pt) node[above=2pt] {1};     

  \draw (0,0) -- (2,0) node[midway, below] {2};           
  \draw (0,0) -- (1,1.5);                                 
  \draw (2,0) -- (1,1.5);                                 

\end{tikzpicture}
    \caption{Enhanced cone in the Higgs branch of Sp(2) at infinite coupling with $N_f=2$.}
    \label{fig:Nf=2enhancedcone}
\end{figure}
\noindent intersecting at the origin. The global symmetry is now $SO(4)\times U(1)$, with the perturbative HS up to order 6 being:
\begin{equation}
\begin{array}{ll}
   \text{HS}(a,b;t)=  &  1+(1+\mu_1^2+\mu_2^2)t^2+ \\
   & +(q+q^{-1})\mu_2t^3+ \\
     & +(1+\mu_2^2+\mu_2^4+\mu_1^4)t^4+ \\
     & +(q+q^{-1})(\mu_2+\mu_2^3)t^5+\\
     & +(1+\mu_2^4+\mu_2^6+\mu_1^6+(q^2+q^{-2}+1)\mu_2^2)t^6+\dots
\end{array}
\end{equation}
and the PL up to order 6 being:
\begin{equation}
\begin{array}{ll}
   \text{PL[HS]}(a,b;t)=  & (1+\mu_1^2+\mu_2^2)t^2+ \\
   & +(q+q^{-1})\mu_2t^3+\\
     & -(2+\mu_1^2+\mu_1^2\mu_2^2)t^4+ \\
     & -(q+q^{-1})(\mu_2+\mu_1^2\mu_2)t^5+ \\
     & +(2\mu_1^2+3\mu_1^2\mu_2^2)t^6+\dots
\end{array}
\end{equation}
notice that the second Dynkin label corresponds to the SU(2) in the enhanced cone, while the first one is the classical SU(2). The generators can be seen to be:
\begin{table}[H]
    \centering
    \begin{tabular}{|c|c|c|c|} \hline
       Generators  & R charge & $SO(4)$ & U(1) \\ \hline
        $M_{\alpha\beta}^+$ & 2 & $\mu_2^2$ & 0 \\
        $M_{\dot{\alpha}\dot{\beta}}^-$ & 2 & $\mu_1^2$ & 0 \\ 
        $S$ & 2 & $1$ & 0 \\
        $I_{\alpha},\tilde{I}_{\alpha}$ & 3 & $\mu_2$ & $+1,-1$ \\
        \hline
    \end{tabular}
    \caption{Infinite coupling generators for $N_f=2$.}
    \label{tab:infinite coupling generators Nf=2}
\end{table}
\noindent Given these generators, we can build 3 singlets at order $t^4$, namely, besides the above mentioned two, also the $S^2$. Therefore in the two singlet constraints at order $4$, we could have $S^2$ entering any one of them, as long as we start with three invariants and only one of them is independent. The constraints are:
\begin{table}[H]
    \centering
    \begin{tabular}{|c|c|c|c|} \hline
      Constraints  & R charge & $SO(4)$ & U(1) \\ \hline
    $\text{tr}M_-^2=0$ & 4 & $1$ & 0 \\
    $\text{tr}M_+^2+S^2=0$ & 4 & $1$ & 0 \\
$M_{\alpha\beta}^+M_{\dot{\gamma}\dot{\delta}}^-=0$ & 4 & $\mu_1^2\mu_2^2$ & 0 \\
     $SM_{\dot{\alpha}\dot{\beta}}^-=0$ & 4& $\mu_1^2$ & 0 \\ \hline
     $ SI_{\alpha}+I_{\beta}M_{\alpha\beta}^+=0$ & 5 & $\mu_2$ & 1 \\
     $ S\tilde{I}_{\alpha}+\tilde{I}_{\beta}M_{\alpha\beta}^+=0$ & 5 & $\mu_2$ & $-1$ \\
      $I_{\gamma}
     M_{\dot{\alpha}\dot{\beta}}^-=0$ & 5 & $\mu_1^2\mu_2$ & +1\\
      $\tilde{I}_{\gamma}
       M_{\dot{\alpha}\dot{\beta}}^-=0$ & 5 & $\mu_1^2\mu_2$ & $-1$\\
      \hline
      $ S^3+ I_{\alpha}\tilde{I}_{\beta}\epsilon_{\alpha\beta}=0$ & 6& $1$ & 0  \\ \hline
    \end{tabular}
    \caption{Infinite coupling constraints for $N_f=2$. We have chosen $M_{\dot{\alpha}\dot{\beta}}^-$ to be associated to the classical cone.}
    \label{tab:constraintsNf=2infinitecoupling}
\end{table}
\noindent The last equation in Tab. \ref{tab:constraintsNf=2infinitecoupling} is a generalized Klein singularity, as the instantons are now vectors. $M_{\dot{\alpha}\dot{\beta}}^-$ represents the classical, non-enhanced cone, hence it is unaffected by $S$ or $I,\tilde{I}$. Indeed, the cone spanned by $M_{\dot{\alpha}\dot{\beta}}^-$ is orthogonal to the cone spanned by all other invariants, as seen from the constraints. A visual representation of the constraints is the following:
\begin{figure}[H]
    \centering

\usetikzlibrary{decorations.markings}

\begin{tikzpicture}[scale=1.2]

\coordinate (O) at (0,0);

\coordinate (L1) at (-2,1);
\coordinate (L2) at (-2,-1);

\draw[thick] (O) -- (L1);
\draw[thick] (O) -- (L2);

\draw[thick, rotate around={0:(-2,0)}] (-2,0) ellipse (0.3 and 1);

\coordinate (R1) at (2,1.6);
\coordinate (R2) at (2,-1.6);

\draw[thick] (O) -- (R1);
\draw[thick] (O) -- (R2);

\draw[thick, rotate around={0:(2,0)}] (2,0) ellipse (0.3 and 1.6);

\node[align=center] at (-4,0) {\textbf{Classical cone}: \\ $\text{tr}M_-^2=0$};
\node[align=center] at (4,0) {\textbf{Enhanced cone}: \\ $\text{tr}M^2_+=S^2$ \\ $SI+MI=0$ \\ $S^3=I \tilde{I}$};

\node[align=center] at (0,-2) {\textbf{Orthogonality}: \\ $M_-S=0 $\\ $M_-I=0 $\\ $M_-\tilde{I}=0$\\$M_-M_+=0$};

\end{tikzpicture}

    \caption{Visual representation of the moduli space of vacua at infinite coupling. The notation of the equations is only schematic, the exact constraints are in Tab. \ref{tab:constraintsNf=2infinitecoupling}.}
    \label{fig:placeholder}
\end{figure}
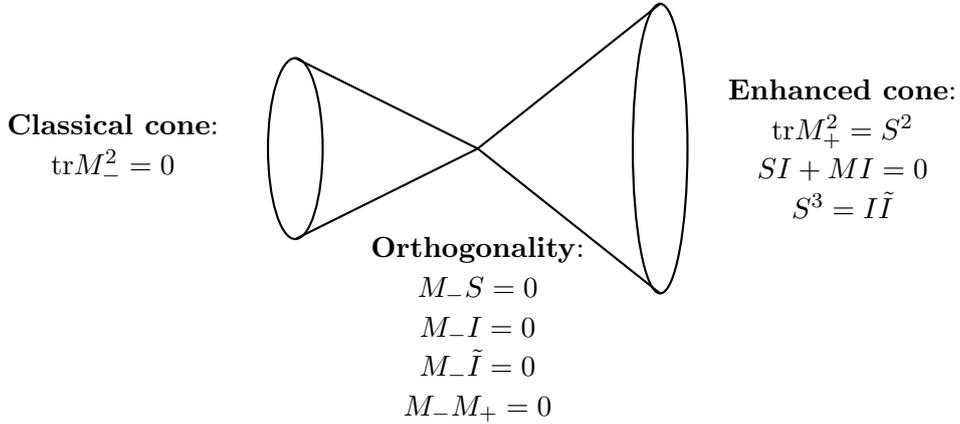
 
\noindent  Again, we can recover the full set of finite coupling relations by putting to zero the instantons in Tab. \ref{tab:constraintsNf=2infinitecoupling}. In turn, we can flow to zero coupling by putting to zero the gaugino bilinear.

\paragraph{Determining the coefficients of the constraints}
We apply the known procedure: we consider the enhanced cone in Fig. \ref{fig:Nf=2enhancedcone} as a 3d Coulomb branch of the quiver gauge theory depicted in the diagram. It is easy to check that such theory is mirror self-dual, hence its Higgs branch is the same as the Coulomb branch. \\
Let us specify the charges under $U(1)_L, U(1)_T$ (the left and top gauge nodes in Fig. \ref{fig:Nf=2enhancedcone}), $U(2)_F$ the global symmetry we obtain once unframing the right node, and $U(1)_F$ for the remaining flavour symmetry. The fields are reported in the following table:
\begin{table}[H]
    \centering
    \begin{tabular}{|c|c|c|c|c|} \hline
    & $U(1)_L$ & $U(1)_T$ & $SU(2)_F$ & $U(1)_F$ \\ \hline
       $A$  & +1 & 0 & $\Box$ & 0 \\
        $\tilde{A}$ & $-1$ & 0 & $\bar{\Box}$ & 0  \\
        $B$ & $-1$ & +1 & 0 & 0\\
        $\tilde{B}$ & +1 & $-1$ & 0 & 0 \\
        $C$ & 0 & $-1$ & 0 & +1 \\
        $\tilde{C}$ & 0 & +1 & 0 & $-1$\\  \hline
    \end{tabular}
    \caption{Matter fields of the 3d quiver gauge theory in Fig. \ref{fig:Nf=2enhancedcone}.}
    \label{tab:fieldsU(1)U(1)}
\end{table}
\noindent Therefore there exist four possible invariants:
\begin{table}[H]
    \centering
    \begin{tabular}{|c|c|} \hline
        Invariant & Definition \\ \hline
        $M_i^j$ & $A_i\tilde{A}^j-A_k\tilde{A}^k\delta_i^j/2$\\
        $S$ & $C \tilde{C}$ \\
        $I$ & $CBA$ \\
        $\tilde{I}$ & $\tilde{A}\tilde{B}\tilde{C}$ \\ \hline
    \end{tabular}
    \caption{Invariants of the Higgs branch of the quiver gauge theory in Fig. \ref{fig:Nf=2enhancedcone}.}
    \label{tab:invariantNf=2}
\end{table}
\noindent We can also write the superpotential of the theory: 
\begin{equation}
    W=\tilde{A}\phi_1 A +B\phi_1 \tilde{B}+\tilde{B}\phi_2B+C\phi_2 \tilde{C}
\end{equation}
which gives rise to the following F-term relations:
\[
\begin{array}{cc}
    A\tilde{A}+\tilde{B}B=0   \\
     B\tilde{B}+\tilde{C}C=0 \\ 
\end{array}
\]
which, written in terms of the invariants, gives:
\[
\tilde{A}^i A_i=A_i \tilde{A}^i=S
\]
thus proving:
\[
\text{tr}M^2=M_i^jM_j^i=(A_i \tilde{A}^j-A_k\tilde{A}^k \delta_i^j/2)( A_j \tilde{A}^i-A_k\tilde{A}^k \delta_j^i/2)=S^2/2
\]

\subsection*{$N_f=1$}
The theory at infinite coupling has a Higgs branch represented by the following magnetic quiver:
\begin{figure}[H]
    \centering
    \begin{tikzpicture}

  \filldraw (0,0) circle (2pt) node[below=2pt] {1};
  \filldraw (2,0) circle (2pt) node[below=2pt] {1};

  \draw (0,0) -- (2,0) node[midway, above] {3};

\end{tikzpicture}
    \caption{Higgs branch of Sp(2) at infinite coupling with $N_f=1$.}
    \label{fig:enter-label}
\end{figure}
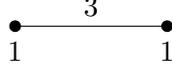
\noindent which represents $\mathbb{C}^2/\mathbb{Z}_3$, i.e. $\mathbf{A}_2$ singularity. The global symmetry is U(1), with the Hilbert series being:
\begin{equation}
    \text{HS}(q,t)=\dfrac{1+t^2+t^4}{(1-q^{-1}t^3)(1-q t^3)} \ ,
\end{equation}
or perturbatively:
\begin{equation}
    \text{HS}(q,t)=1+t^2+(q+q^{-1})t^3+\dots+(q^2+1+q^{-2})t^6+\dots
\end{equation}
and the PL being:
\begin{equation}
    \text{PL}(q,t)=t^2+t^3(q+q^{-1})-t^6 \ .
\end{equation}
By comparing the HS and the PL we are able to conclude the existence of the following generators of the Higgs branch chiral ring:
  \begin{table}[H]
        \centering
        \begin{tabular}{|c|c|c|} \hline
            Generators & R-charge & U(1) \\ \hline
        X & 2 & 0 \\
            $I,\tilde{I}$ & 3 & $ 1,-1$ \\ \hline
        \end{tabular}
        \caption{Generators for Sp(2) with $N_f=2$ at infinite coupling.}
        \label{tab:Nf=1 generators}
    \end{table}
\noindent where $X$ is a linear combination of the gaugino bilinear $S$ and the unique meson $M$. The only constraint is:
\[
X^3+I\tilde{I}=0 \ .
\]
where $X=a M+bS$. The coefficients $a,b$ are unknown. However, out of consistency with the higher flavour cases, we expect:
\begin{equation}
    M^2=S^2 \ ,
    \label{eq:msquared}
\end{equation}
whence we conclude:
\[
S^3 + I \tilde{I}=0 \ .
\]
At finite coupling, the only constraint is Eq. \ref{eq:msquared}, which at zero coupling reproduces the existence of a so-called fat point, described by:
\[
M^2=0 \ .
\]

\subsection*{Pure theory}
\paragraph{Infinite coupling} The theory at infinite coupling has the same quiver representation, HS and PL as the $N_f=1$ case, with the difference that $X$ is now undeniably equal to the gaugino bilinear, as there are mesons. Hence the unique constraint is:
    \begin{table}[H]
        \centering
        \begin{tabular}{|c|c|c|} \hline
            Constraint & R-charge & U(1) \\ \hline
            $aS^3 +bI\tilde{I}=0$ & 6  & 0\\ \hline
        \end{tabular}
        \caption{Constraint for pure Sp(2) at infinite coupling.}
        \label{tab:pure constraint}
    \end{table}
\noindent Notice that the coefficients $a,b$ are known to be nonzero, since the moduli space is $\mathbb{C}^2/\mathbb{Z}_3$, hence they can be reabsorbed in the definitions of the invariants.

\subsection{Flavours with symmetry enhancement}\label{sec:symmetryenhancement}
For $Sp(2)$, the two cases with symmetry enhancement are $N_f=8$ and $N_f=9$.

\subsection*{$N_f=8$}
\paragraph{Infinite coupling}
The Higgs branch at infinite coupling has global symmetry $SO(16)\times SU(2)$ and is represented by the following quiver belonging to the $E_7$ exceptional family:
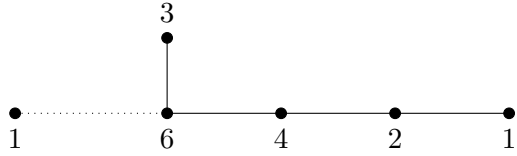
\begin{figure}[H]
    \centering
   \begin{tikzpicture}

  \filldraw (0,-1) circle (2pt) node[below=2pt] {1};
  \filldraw (2,-1) circle (2pt) node[below=2pt] {$6$};
  \filldraw (5,-1) circle (2pt) node[below=2pt] {2};
  \filldraw (2,0) circle (2pt) node[above=2pt] {3};
  \filldraw (6.5,-1) circle (2pt) node[below=2pt] {1};
  \filldraw (3.5,-1) circle (2pt) node[below=2pt] {4};

  \draw[dotted] (0,-1) -- (2,-1);
  \draw (2,-1) -- (2,0);
  \draw (2,-1) -- (3.5,-1);
  \draw (3.5,-1) -- (5,-1) -- (6.5,-1);

\end{tikzpicture}
    \caption{Quiver diagram for the infinite coupling Higgs branch for $N_f=8$}
    \label{fig:infinitecouplingNf=7}
\end{figure}
\noindent The infinite coupling Higgs branch has the following\footnote{The fugacity $\nu$ is related to the SU(2) symmetry, while the $\mu_i$'s represent SO(16).} HS:
\begin{equation}
   \begin{array}{ll}
       \text{HS}(t) & =1+(\nu^2\cdot 1+1 \cdot \mu_2)t^2+ \\
        & +\nu \mu_8 t^3+\\
        & +(1\cdot 1+\dots)t^4+\\
        & +\nu (\mu_8+\mu_2+\dots)t^5+\\
        & +(\nu^2\cdot 1+\nu^2\mu_4+1 \cdot \mu_6+1 \cdot \mu_2+\dots)t^6+\dots \\
   \end{array}
\end{equation}
whose PL is:
\begin{equation}
    \begin{array}{ll}
       \text{PL[HS]}(t) & =1+(\nu^2 \cdot 1+1 \cdot \mu_2)t^2+ \\
        & +\nu \mu_8t^3+\\
        & -(1\cdot 1+1 \cdot \mu_1^2)t^4+\\
        & -\nu (\mu_8+\mu_1\mu_7)t^5+\\
        & -(1 \cdot (\mu_2+\mu_6)+\nu^2(\mu_4+1))t^6+\dots \\
    \end{array}
\end{equation}
We recall that the spinor representation of $D_8$ is real and its indices can be contracted with:
\[
\delta_{\alpha \beta} \quad \delta_{\dot{\alpha}\dot{\beta}} \quad (\gamma_i)_{\alpha\dot{\beta}}\quad (\gamma_{ij})_{\alpha\beta}
\]
The generators are:
\begin{table}[H]
    \centering
    \begin{tabular}{|c|c|c|c|}\hline
      Generators   &  R-charge & $SO(16)$ & $SU(2)$ \\ \hline
       $M_{ij}$  & 2 & $\mu_2$ & $1$ \\
       $S_{(ab)}$ & 2 & $1$ & $\nu^2$  \\ \hline
       $I_{a}^{\alpha}$ & 3 & $\mu_8$ & $\nu$ \\ \hline 
    \end{tabular}
    \caption{Generators for the infinite coupling Higgs branch chiral ring at $N_f=8$.}
    \label{tab:Nf8generators}
\end{table}
\noindent where $S^{(ab)}$ is composed of the gaugino bilinear and two instantons of charge $\pm2$, transforming in the singlet of $SO(16)$. The constraints are:
\begin{table}[H]
    \centering
    \begin{tabular}{|c|c|c|c|} \hline
       Constraints  & R-charge & $SO(16)$ & $SU(2)$ \\ \hline
        $M_{ij}M_{kl}\delta_{jk}=\text{tr}S^2 \delta_{ij}$ & 4 & $\mu_1^2+1$ & $1$ \\ \hline
        $S_{cd}I^{\alpha}_e\epsilon^{ed}(\gamma_j)_{\alpha \dot{\beta}}+M_{ij}(\gamma_i)_{\rho\dot{\beta}}I^{\rho}_c=0$ & 5 & $\mu_1\mu_7+\mu_8$ & $\nu$ \\ \hline
    $S_{ab}^3+I^{\alpha}_{(a}I^{\beta}_{b)}\delta_{\alpha\beta}=0$ & 6 & $1$ & $\nu^2$\\
    $S_{ab}M_{ij}M_{kl}+I^{\alpha}_aI^{\beta}_b(\gamma_{ijkl})_{\alpha\beta}=0$ & 6 & $\mu_4$ & $\nu^2$ \\
    $\text{tr}S^2M_{ij}+I^{\alpha}_aI^{\beta}_b(\gamma_{ij})_{\alpha\beta}\epsilon^{ab}=0$ & 6 & $\mu_2$  & $1$ \\
    $M_{ij}M_{kl}M_{mn}+ I^{\alpha}_aI^{\beta}_b(\gamma_{ijklmn})_{\alpha\beta}\epsilon^{ab}=0$ & 6 & $\mu_6$ & $1$ \\ \hline
    \end{tabular}
    \caption{Infinite coupling constraints for $N_f=8$}
    \label{tab:constraintsNf=8}
\end{table}
\noindent Although the form of the constraints is different from Tab. \ref{tab:infinitecouplingconstraintsGENERAL} due to the presence of SU(2) representations, the logic is the same. Indeed we have the Casimir correction, the eigenvalue equation and the instanton bilinear correction. We have no Hodge correction since we are outside the bound $N_f \leq 2k+1$. The instanton bilinear correction is hidden by the representations of SU(2).

\subsection*{$N_f=9$}

\paragraph{Infinite coupling}
The global symmetry on the Higgs branch is now $SO(20)$, enhancing the classical $SO(18)\times U(1)$, with the chiral ring represented by the following quiver belonging to the $E_8$ exceptional family:
\begin{figure}[H]
    \centering
   \begin{tikzpicture}

  \filldraw (0,-1) circle (2pt) node[below=2pt] {1};
  \filldraw (2,-1) circle (2pt) node[below=2pt] {8};
  \filldraw (5,-1) circle (2pt) node[below=2pt] {2};
  \filldraw (2,0) circle (2pt) node[above=2pt] {4};
  \filldraw (3.5,-1) circle (2pt) node[below=2pt] {5};

  \draw[dotted] (0,-1) -- (2,-1);
  \draw (2,-1) -- (2,0);
  \draw (2,-1) -- (3.5,-1);
  \draw (3.5,-1) -- (5,-1);

\end{tikzpicture}
    \caption{Magnetic quiver for the infinite coupling Higgs branch with $N_f=9$.}
    \label{fig:infinitecouplingNf=9}
\end{figure}
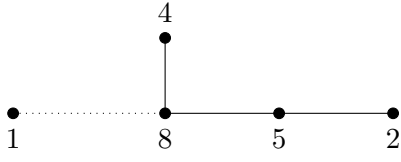
\noindent We recall that the spinor representation of $SO(20)$ is pseudoreal and its indices can be contracted with:
\[
\epsilon_{\alpha\beta} \quad (\gamma_i)_{\alpha\dot{\alpha}} \quad (\gamma_{ij})_{\alpha\beta} \quad \text{etc.}
\]
The HS of the chiral ring is:
\begin{equation}
    \begin{array}{ll}
        \text{HS}(t) & = 1+\mu_2t^2+\\
         & +\mu_{10}t^3+\\
         & +(1+\mu_4+\mu_2^2)t^4+\\
         & + (\mu_{10}+\mu_2\mu_{10})t^5+\\
         & +(\mu_{10}^2+\mu_2+\mu_2\mu_4+\mu_2^3+\mu_6)t^6+\dots \\
    \end{array}
\end{equation}
whose PL is:
\begin{equation}
    \begin{array}{ll}
       \text{PL[HS]}(t)  & =  1+\mu_2t^2 +\mu_{10}t^3+\\
         & -\mu_1^2t^4+\\
         & -\mu_1\mu_8t^5+\\
         & -(\mu_2+\mu_6-\mu_1^2)t^6+\dots
    \end{array}
\end{equation}
where we recognize the generators:
\begin{table}[H]
    \centering
    \begin{tabular}{|c|c|c|} \hline
       Generators  & R-charge & $SO(20)$  \\ \hline
        $M_{ij}$ & 2 & $\mu_2$ \\ \hline
        $I_{\alpha}$ & 3 & $\mu_{10}$ \\ \hline
    \end{tabular}
    \caption{Infinite coupling generators for Nf=9}
    \label{tab:infinitecouplingNf=9}
\end{table}
\noindent And the constraints:
\begin{table}[H]
    \centering
    \begin{tabular}{|c|c|c|}\hline
        Constraints & R-charge & $SO(20)$  \\ \hline
        $M_{ij}M_{jl}-\delta_{il}\text{tr}M^2/20=0$ & 4 & $\mu_1^2$ \\ \hline
        $M_{ij}(\gamma_{j})_{\dot{\alpha}\beta}I_{\rho}\epsilon^{\beta\rho}-M_{kl}(\gamma_{ikl})_{\dot{\alpha}\beta}I_{\rho}\epsilon^{\beta\rho}/20=0$ & 5 & $\mu_1\mu_9$ \\ \hline
       $M_{ij}\text{tr}M^2+I_{\alpha}I_{\beta}(\gamma_{ij})^{\alpha\beta}=0$ & 6 & $\mu_2$  \\
       $M_{[ij}M_{kl}M_{mn]}+I_{\alpha}I_{\beta}(\gamma_{ijklmn})^{\alpha\beta}=0$ & 6 & $\mu_6$ \\ \hline
    \end{tabular}
    \caption{Infinite coupling constraints for $N_f=9$}
    \label{tab:Nf=9infcouplingconstraints}
\end{table}
\noindent These constraints, which agree with the one found in \cite{hanany2018small}, obey the same rules as Tab. \ref{tab:infinitecouplingconstraintsGENERAL}, except now there are only two generators. The larger symmetry group simplifies the form of the constraints and synthetizes them in just four equations: the (absence) of a Casimir correction, the eigenvalue equation, and the instanton bilinear corrections. If we were to branch $SO(20)\to SO(18)\times U(1)$, there would be also additional constraints involving instantons of charge $\pm 2$, which are now hidden in the mesons of Tab. \ref{tab:infinitecouplingNf=9}.

\section{Conclusions}
We computed the chiral ring at infinite coupling for $Sp(k)$ theories and we gave a consistent explanation and classification of the constraints between the generators, summarized in Tab. \ref{tab:infinitecouplingconstraintsGENERAL}. We considered the RG flow activated by the instanton mass, and we showed that the resulting chiral ring is different from the mere F-term equations. The difference consists in the gaugino bilinear, which is a nilpotent operator at weak coupling. Such nilpotency implies the existence of a discrete set of operators in the chiral ring, which was previously unaccounted for, labeled by powers of $S$, besides the continuous variety. \\

\noindent In addition to the incompleteness of the F-term constraint in describing the entire moduli space of vacua, we also showed that the Higgs branch is perturbatively corrected due to the gaugino bilinear itself. This result clashes with non renormalization theorems developed \cite{seiberg1993naturalness} in 4d $\mathcal{N}=1$, altough it is not entirely clear what assumptions of the theorem fail in the presence of a nilpotent operator. As mentioned previously, it would be interesting to reproduce these results at weak coupling by means of some perturbative computation, perhaps along the lines of \cite{cachazo2003chiral, argurio2004introduction}. It would be furthermore interesting to truncate to 4d, with same amount of supersymmetry or less, and compare with the known results in such dimension.\\

\noindent It would be helpful to find physically intuitive ways to describe the consequences of the corrections at finite and infinite coupling, along the lines of 4d $\mathcal{N}=1$ with $k=N_c$, where the origin of the moduli space is made non-singular by nonperturbative effects. It would also be of interest to adapt the reasoning performed in this paper to more complicated theories, which feature CS levels (like $SU(k)$ theories) or even baryons (like $SO(k)$ and $SU(k)$ gauge theories). It could be possible that the degree of nilpotency of the gaugino bilinear is affected by the level itself.  Furthermore, it would be interesting to detect why the left procedure of Tab. \ref{tab:proceduresdonotcommute} does not provide the full set of generators and relations. Namely, it would be interesting to read the existence and degree of nilpotency of the gaugino bilinear out of the finite coupling braneweb, hence improving the technique of \cite{cabrera2019tropical}.

\section*{Acknowledgments}
E.VdD. would like to thank Riccardo Argurio, Sam Bennett, Stefano Cremonesi and Noppadol Mekareeya for interesting comments. E.VdD. and A.H. gratefully acknowledge the hospitality of the Simons center for Geometry and Physics during the 2025 Simons summer workshop. The work of A.H. and E.VdD. is partially supported by STFC Consolidated Grant ST/X000575/1. 

\bibliographystyle{JHEP}
\bibliography{bibli.bib}

\end{document}